\documentclass[fleqn,usenatbib]{mnras}

\usepackage{newtxtext,newtxmath}

\usepackage[T1]{fontenc}
\usepackage{url}

\DeclareRobustCommand{\VAN}[3]{#2}
\let\VANthebibliography\thebibliography
\def\thebibliography{\DeclareRobustCommand{\VAN}[3]{##3}\VANthebibliography}

\usepackage{graphicx}	
\usepackage{amsmath}	
\usepackage{comment}

\usepackage[subrefformat=parens]{subcaption}

\usepackage{multirow}

\newcommand{\mpc}{$h^{-1}$Mpc}
\newcommand{\gpc}{$h^{-1}$Gpc}
\newcommand{\ave}[1]{\left\langle{#1}\right\rangle}

\title[Photometric BAO]{Robustness of Baryon Acoustic Oscillations Measurements with Photometric Redshift Uncertainties}

\author[K. Ishikawa et al.]{
Keitaro Ishikawa,$^{1}$\thanks{E-mail:ishikawa.keitaro.r3@s.mail.nagoya-u.ac.jp}
Tomomi Sunayama,$^{2,3}$
Atsushi J. Nishizawa,$^{4,5,2}$
Hironao Miyatake,$^{2,5,6}$\
\newauthor
and
Takahiro Nishimichi$^{7,8,6}$
\\
$^{1}$Physics Department, Nagoya University, Furocho, Chikusa, Nagoya, Aichi, 464-8602, Japan\\
$^{2}$Kobayashi Maskawa Institute, Nagoya University, Furocho, Chikusa, Nagoya, Aichi, 464-8602, Japan\\
$^{3}$Department of Astronomy and Steward Observatory, University of Arizona, 933 N Cherry Ave, Tucson, AZ, 85719, USA\\
$^{4}$DX Center, Gifu Shotoku Gakuen University, 1-1 TakakuwaNishi, Yanaizucho, Gifu, Gifu, 501-6194, Japan\\
$^{5}$Institute for Advanced Research, Nagoya University, Furocho, Chikusa, Nagoya, Aichi, 464-8602, Japan\\
$^{6}$Kavli Institute for the Physics and Mathematics of the Universe (WPI), The University of Tokyo Institutes for Advanced Study (UTIAS), The University of Tokyo, Chiba 277-8583, Japan\\
$^{7}$Department of Astrophysics and Atmospheric Sciences, Faculty of Science,
Kyoto Sangyo University, Motoyama, Kamigamo, Kita-ku, Kyoto 603-8555, Japan\\
$^{8}$Center for Gravitational Physics and Quantum Information, Yukawa Institute for Theoretical Physics, Kyoto University, Kyoto 606-8502, Japan
}

\date{Accepted XXX. Received YYY; in original form ZZZ}

\pubyear{2015}

\begin{document}
\label{firstpage}
\pagerange{\pageref{firstpage}--\pageref{lastpage}}
\maketitle

\begin{abstract}
We investigate the robustness of baryon acoustic oscillations (BAO) measurements with a photometric galaxy sample using mock galaxy catalogues with various sizes of photometric redshift (photo-$z$) uncertainties.
We first conduct the robustness of BAO measurements, assuming we have a perfect knowledge of photo-$z$ uncertainties. We find that the BAO shift parameter $\alpha$ can be constrained in an unbiased manner even for 3\% photometric redshift uncertainties up to $z\sim 1$.
For instance, $\alpha=1.006 \pm 0.078$ with 95\% confidence level is obtained from 3\% photo-$z$ uncertainty data at $z=1.03$ using the sample of  $M_* \ge 10^{10.25} M_{\sun}/h^2$.
We also find that a sparse galaxy sample,
e.g. $<2\times10^{-4}$ [$h$ Mpc$^{-1}]^3$
causes additional noise in the covariance matrix calculation and can bias the constraint on $\alpha$. 
Following this, we look into the scenario where incorrect photometric redshift uncertainties are assumed in the fitting model. We find that underestimating the photo-$z$ uncertainty leads to a degradation in the constraining power on $\alpha$.
However, the constrained value of $\alpha$ is not biased.
We also quantify the constraining power on $\Omega_{\rm m0}$ assuming the LSST-like covariance and find that the 95\% confidence level is $\sigma(\Omega_{\rm m0})\sim0.03$-$0.05$ corresponding to the photo-$z$ uncertainties of 1\% to 3\% respectively.
Finally, we examine whether the skewness in the photometric redshift can bias the constraint on $\alpha$ and 
confirm that the constraint on $\alpha$ is unbiased, even assuming a Gaussian photo-$z$ uncertainty in our model.
\end{abstract}




\section{Introduction}
\label{sec:introduction}
Baryon acoustic oscillations (BAOs) are a signature of the early universe imprinted in the distribution of galaxies and dark matters at the present epoch. 
As the early universe expands and its temperature drops, sound waves propagating in the baryon-photon plasma freeze into the matter distribution at decoupling. 
This leaves a characteristic peak in two-point correlation functions where the position of the peak is about 150~Mpc. 
The BAO method utilises this characteristic scale of the sound horizon as a standard ruler and enables us to constrain cosmological parameters such as $\Omega_{\rm m0}$, $H_0$, and the dark-energy equation-of-state parameter, $w$ \citep[for details, see a review by][]{Weinberg+2013}.
In addition, the BAO method is the least sensitive to astrophysical uncertainties and one of the most powerful cosmological probes since the first detection with a spectroscopic galaxy survey by Sloan Digital Sky Survey \citep[SDSS;][]{Eisenstein+2005}. 

The imprint of BAO appears as a peak in a galaxy clustering signal in real space or as a wiggle in Fourier space. 
Spectroscopic survey data have an advantage in the BAO method as its accurate redshift information enables us to measure the three-dimensional correlation functions \citep{Totsuji&Kihara1969,Groth&Peebles1977}.

While spectroscopic galaxy surveys provide a three-dimensional map of the Universe, photometric galaxy surveys provide images of galaxies and the two-dimensional positions of these galaxies. The redshift of these galaxies is inferred from the broadband filters and is known to have larger uncertainties than the spectroscopic redshifts. 
The advantage of photometric surveys is that they can collect a large survey area more efficiently with deep magnitude, which yields a higher number density of galaxies, and the BAO peak can be measured using a galaxy sample from photometric survey data.

Current and future galaxy surveys such as the Hyper Suprime-Cam (HSC) survey \citep{Aihara+2018}, the Dark Energy Survey\footnote{\url{ https://www.darkenergysurvey.org}} (DES) \citep{DES2005}, the Kilo Degree Survey\footnote{\url{http://kids.strw.leidenuniv.nl/}} (KiDS) \citep{KiDs2015}, the Rubin Observatory Legacy Survey of Space and Time\footnote{\url{https://www.lsst.org}} (LSST) \citep{LSST2009}, \textit{Euclid}\footnote{\url{ https://sci.esa.int/web/euclid}} \citep{euclid2018}, and the Nancy Grace Roman Telescope\footnote{\url{https://wfirst.gsfc.nasa.gov}} \citep{WFIRST2019} aim to probe the physical nature of dark energy. 
The advantage of 
using a photometric galaxy sample is that 1) the number density of galaxies can be larger than a spectroscopic galaxy sample, which potentially reduces shot noise, and 2) the BAO can be measured in an area of the sky where a spectroscopic galaxy sample does not exist, such as the southern hemisphere sky covered by DES and LSST.
    
\cite{Hutsi2010} first measures the photometric BAO using maxBCG cluster sample from SDSS using a 3D power spectrum and \cite{Ho:2012, Seo+2012} follows using the SDSS photometric galaxy sample where they measure a projected correlation function in harmonic space. \cite{Abbott+2019} uses the DES-Y1 photometric galaxy sample and measures the projected correlation function as a function of the angular separation and the comoving transverse separation, and extracts the BAO scale using a proposed template by \cite{Ross+2017b}. \cite{Abbott+2022} uses the DES-Y3 data and measures the projected correlation function as a function of the angular separation in the harmonic space. 
Due to its large redshift uncertainties, projected correlation functions are commonly used to measure the BAO peak from the photometric galaxy surveys.

However, several papers have explored the measurement of the BAO peak using three-dimensional correlation functions from the photometric surveys since the photometric redshift estimate will be improved and photometric redshift uncertainties will be reduced for future photometric surveys like LSST. This can potentially enable us to extract more information than the projected correlation functions.
\cite{Ross+2017b} studies the 3D two-point correlation function incorporating the photo-$z$ effect and the dependence on the cosine of the angle to the line-of-sight, $\mu$ in 3D and projected correlation functions.
They show that the BAO peak is smeared by photo-$z$ error, especially near the line-of-sight direction. Therefore, they measure the BAO using the correlation functions with the radial separation perpendicular to the line-of-sight direction.
\cite{Chaves-Montero+2018} also studies the effect of the sub-per cent photo-$z$ error on the BAO using multipole power spectra, particularly focusing on small-scale non-linearity.
\cite{Chan+2022b} quantify the impact of photo-$z$ uncertainties on the three-dimensional correlation function and demonstrate that the location of the BAO peak can be measured using the three-dimensional correlation function from the DES-Y3 data \citep{Chan+2022a}. 
This paper aims to perform complementary studies on the impact of the photometric redshift uncertainties and their biased estimates on the BAO method. Specifically, we investigate the effects of photometric redshift uncertainties on the measurement of the BAO peak using three-dimensional two-point correlation functions.
\begin{itemize}
\item{We focus on the feasibility of using the 3D correlation function, integrating the redshift space correlation function over the $\mu$ direction. It is well known that the BAO scale in the redshift space correlation function exhibits a strong dependence on $\mu$, however; we will show that it offers an unbiased estimate of the BAO scale once we correctly model the photometric redshift uncertainty.}
\item{We relax the assumption of the non-linear smearing of the BAO peak, parameterised by $\Sigma_{\mathrm{NL}}$. During the optimisation process, we simultaneously fit the parameter $\Sigma_{\mathrm{NL}}$ which is highly correlated with the photometric redshift uncertainty, instead of fixing to its expected value from the mock simulation.
}
\item{In order to validate our method, we consider several different situations. Firstly, we assume that we can have a perfect knowledge of the photo-$z$ uncertainty, and secondly, we fit with a wrong photo-$z$ uncertainty model, enabling us to quantify the importance of a priori knowledge of the photo-$z$ uncertainty. Finally, we repeat the analysis at different redshifts and different population of the sample.}
\item{
Furthermore, we explore the robustness of the photo-$z$ BAO when we do not have a full shape of the distribution function of the photo-$z$ uncertainty.}
\end{itemize}

This paper is organised as follows. \S~\ref{sec:theory} indicates how we construct the template model and show the fitting procedure.
\S~\ref{sec:simulation} introduces the galaxy mock catalogue used to verify the photo-$z$ effects in the model.
\S~\ref{sec:results} is devoted to showing our results. \S~\ref{ssec:bao_measure_perror_isknown} is the result when we have a perfect knowledge of the photo-$z$ uncertainty, while in  \S~\ref{ssec:bao_measure_perror_isunknown},  we show the result when the photo-$z$ uncertainty is unknown, and \S~\ref{ssec:bao_measure_perror_isknown_cross} is for the cross-correlation between spec-$z$ and photo-$z$ samples.
\S~\ref{sec:discussion} discusses whether the parameter to capture the BAO peak can be constrained properly when assuming a different cosmological parameter. In addition, we explore whether the result is affected when the photo-$z$ distribution is skewed non-Gaussian.
We conclude in \S~\ref{sec:summary}.

\section{Theory}
\label{sec:theory}

\subsection{model ingredients}
\label{ssec:model}
In this section, we describe the model of the correlation function in redshift space used as a template to fit to the mock BAO data. We first begin with the real-space expression.
The three-dimensional two-point correlation function in real space is 
\begin{align}
    \xi_{\mathrm{m}}(r) = \int \text{d}k \dfrac{k^2}{2\pi^2} P^{\mathrm{NL}}_{\mathrm{m}}(k) j_0(kr) \label{eq:xi_r}, 
\end{align}
where $j_0$
is the spherical Bessel function, and $P_{\mathrm{m}}^{\text{NL}}(k)$ is the non-linear matter power spectrum \citep{Eisenstein+2007}, 
\begin{align}
    P^{\text{NL}}_{\text{m}}(k) = \left( P_{\text{lin}}(k) - P_{\text{nw}}(k) \right)e^{-k^2 \Sigma_{\text{NL}}^2(z) /2} + P_{\text{nw}}(k) \label{eq:Pk}, 
\end{align}
where $P_{\text{lin}}(k)$ is the linear matter power spectrum, 
and $P_{\text{nw}}(k)$ is the no-wiggle matter power spectrum using the fitting formula of
\cite{Eisenstein_Hu1998}. 
It is well known that the non-linear evolution of dark matter may cause a smoothing effect of the BAO peak in the correlation function. Although the effect can be precisely described by the higher order perturbation \citep[e.g.][]{TaruyaBernardeau:2012} or emulator \citep[e.g.][]{Nishimichi+2019}, here we employ an empirical model of Eq.~\eqref{eq:Pk}, 
keeping the degree of freedom for the amount of smoothing via free parameter $\Sigma_a$,
\begin{align}
    \Sigma_{\text{NL}}(z) = \Sigma_a D(z)/D(0) \label{eq:Sigma_nl}, 
\end{align}
where $D(z)$ is a linear growth factor 
and $\Sigma_a$ is 
a free parameter describing the strength of the smoothing.
The power spectrum of the linear perturbation theory and the growth factor are computed using the publicly available code \texttt{CLASS} \citep{CLASS2011} and fitting formula of \cite{Carroll+1992}, respectively.

\begin{figure}
    \begin{center}
        \includegraphics[width=\linewidth,keepaspectratio]{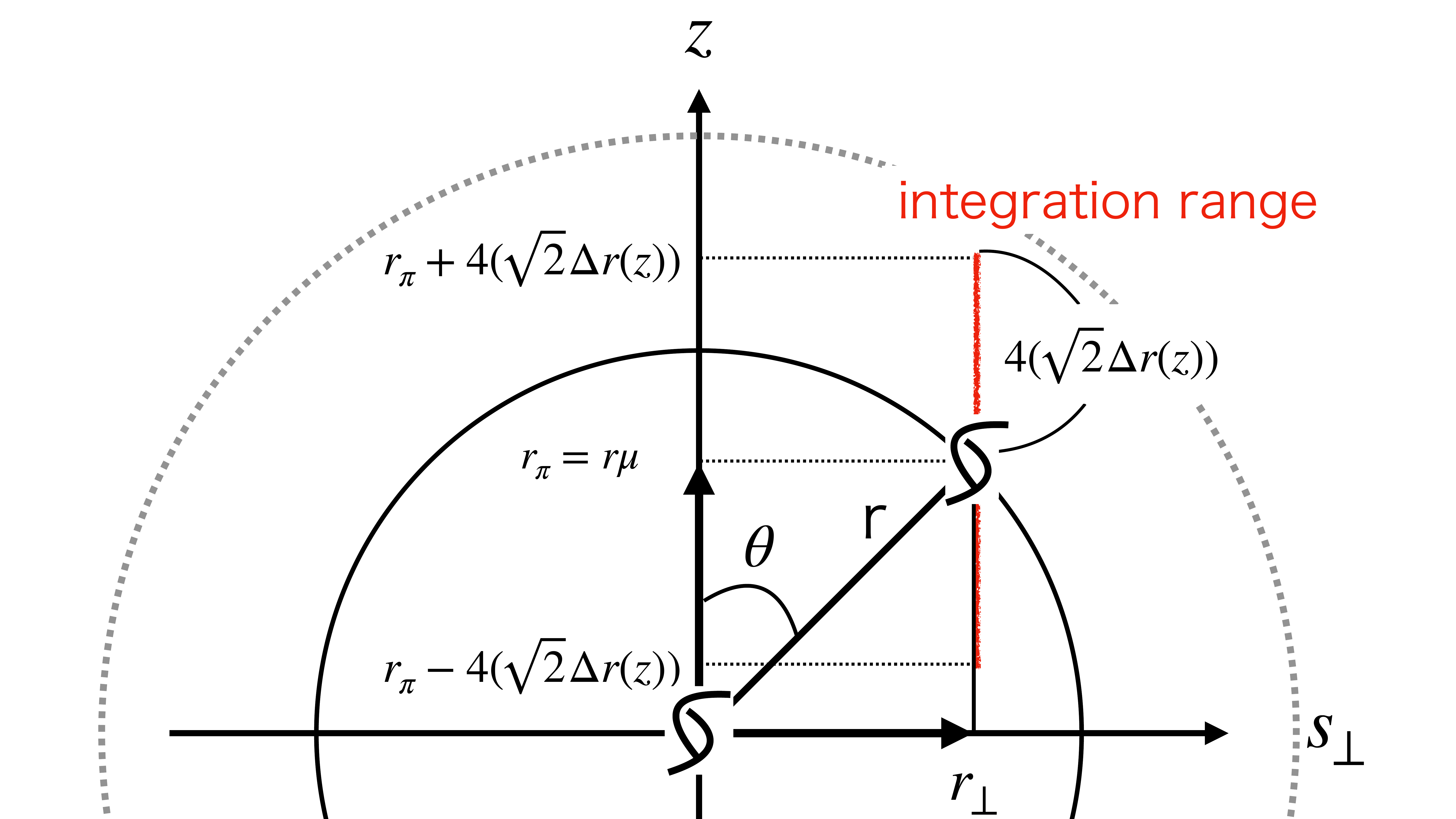}
        \caption{Schematics of our model of three-dimensional correlation function with photo-$z$ uncertainties.
        $4 \sqrt{2}\Delta r(z)$ is the integration range which is sufficient for converging the integral (Eq.~\eqref{eq:xi_int} and the factor $\sqrt{2}$ originates from the effect due to the Gaussian photo-$z$ error of the two galaxies being forced to one side.}
        \label{fig:photoz_conv}
    \end{center}
\end{figure}

In real observation, the position of galaxies is measured in redshift space. Although the magnitude of the current photometric redshift uncertainty is large and the effect of the redshift space distortion might be subdominant, we consider the effect for the complete description.
Within the linear regime, the
correlation function in redshift space
is modulated from the real-space correlation function \citep{Kaiser1987},
\begin{align}
    \xi^{(s)}(r,\mu) = \left(1+\beta (z) \mu^2 \right)^2 \xi_{\mathrm{m}}(r)
    \label{eq:xi_s},
\end{align}
under the plane-parallel approximation,
where $\beta(z)=f(z)/B$, $f$ is a linear growth rate, and $\mu$ is a cosine of galaxy separation and line of sight.
Note that the galaxy correlation function should be multiplied by the square of a linear galaxy bias factor.
We take this into account later in Eq.~\eqref{eq:fitting_formula} 
and parameterise it by constant $B$.
We treat $B$ as the nuisance parameter responsible for the overall amplitude, which is, in the end, marginalised when we extract the shift parameter $\alpha$ detailed in Eq.~\eqref{eq:alpha}.
The model of Eq.~\eqref{eq:xi_s} does not include the uncertainty of the radial position of galaxies. While the model is valid for the spec-$z$ sample, we cannot ignore the effect when the redshift measurement holds a non-negligible uncertainty size. The uncertainty on the radial position can be considered by convolving the probability distribution \citep{Ross+2017b, Chan+2022b}
\begin{align}
    \xi_{\text{P}}^{(s)}(r,\mu) = \int \text{d}r'_{\pi} {\cal P}(r'_{\pi} - r\mu) \xi_{s}(\sqrt{{r'}^2_{\pi} + r^2(1-\mu^2)})
    \label{eq:xi_int}, 
\end{align}
where ${\cal P}(r'_{\pi} - r\mu)$ is a probability distribution function of finding the galaxy pair as a function of radial separation.
For the complete description of the model, here we assume the probability obeys Gaussian distribution, ${\cal N}(0,\sqrt{2}\sigma_P)$, where $\sigma_P$ is a typical photometric redshift uncertainty of the survey assumed.
We generalise the assumption of Gaussian in \S\ref{ssec:skewness}.
The integration range of Eq.~\eqref{eq:xi_int} should be taken so that the integral converges and we find that $\pm 4\sqrt{2}\Delta r(z)$ is sufficient, where $\Delta r$ corresponds to the radial uncertainty due to the photo-$z$ in the unit of \mpc.
Fig.~\ref{fig:photoz_conv} illustrates the configuration of the photo-$z$ uncertainty model.
Finally, we integrate the model over angular direction to obtain the angle averaged correlation function, 
\begin{align}
    \overline{\xi}^{(s)}_{\rm P}(r) = \int^1_0 \! \text{d}\mu \xi^{(s)}_{\rm P}(r,\mu) \label{eq:xi_temp}.
\end{align}

\subsection{fitting procedures}
\label{ssec:fitting_procedures}
Now we compare our model with the mock simulation. We assume that the bias is scale independent, and we further introduce an additive polynomial term to absorb any non-linear effects, including non-linear dark matter clustering or non-linear galaxy (halo) bias \citep{Xu+2012}, 
\begin{align}
    \xi^{\text{fit}}(r) &= B^2 \xi^{(s)}_{\text{P}}(\alpha r|\Sigma_a) + A(r), \label{eq:fitting_formula} \\
    A(r) &= \dfrac{a_1}{r^2} + \dfrac{a_2}{r} + a_3 \label{eq:polynominal},
\end{align}
where $B, a_1, a_2, a_3, \Sigma_a, $ and the shift parameter $\alpha$ are all fitting parameters.
The nuisance parameter $B$ corresponds to the Kaiser factor incorporating the photo-$z$ effect in redshift space, which can be interpreted on one side as the galaxy linear bias but on other hand, a relative amplitude of the correlation function to the empirically introduced polynomial term, $A(r)$.
The shift parameter measures the ratio of the three-dimensional location of the BAO peak in observation to that of the fiducial cosmology, which we need to assume when the galaxy redshifts are converted to the radial comoving distance prior to the measurement of the correlation function. It forms
\begin{align}
    \alpha = \dfrac{[D_V(z)/r_s]_{\text{obs}}}{[D_V(z)/r_s]_{\text{fid}}}, \label{eq:alpha}
\end{align}
where $r_s$ is the sound horizon at the drag epoch i.e. BAO scale and $D_V(z)$ is the dilation scale, that is, 
\begin{align}
    D_V(z) = \left[ D_A(z)^2 \dfrac{cz}{H(z)} \right]^{1/3}, \label{eq:D_V}
\end{align}
where $D_A(z)$ is the angular diameter distance. This equation states that it has the information of the perpendicular direction and that of the parallel one, which is expected to be smeared by the photo-$z$ error.

We fit the model by minimising $\chi^2$
\begin{align}
    \chi^2 = \sum
    \left[ \xi_{\rm mock}(r_1) - \xi^{\rm fit}(r_1) \right]
    {\rm Cov}^{-1}(r_1, r_2)
    \left[ \xi_{\rm mock}(r_2) - \xi^{\rm fit}(r_2) \right]. \label{eq:chi2}
\end{align}
The definition of covariance in Eq.~\eqref{eq:chi2} will be
given in \S~\ref{ssec:measurement}.
The summation runs over $r_1$ and $r_2$ in the fitting range, which we take from
30 \mpc\, to 150 \mpc\ to use the large-scale information and capture the BAO peak.
For the minimisation, we use publicly available Python code, \texttt{scipy.optimize.curve\_fit}\footnote{https://scipy.org} with full covariance and least-square mode.

\section{Simulation}
\label{sec:simulation}

\subsection{Galaxy Mock Catalogue}
\label{ssec:mockcatalog}
    
In this section, we briefly revisit the galaxy mock catalogue. The full description can be found in \cite{Nishimichi+2019} and \cite{ishikawa+2021}.

The mock catalogue is based on the Dark Quest suite of N-body simulations performed using the publicly available code, \texttt{Gadget-2} \citep{Springel2005}. The initial condition is generated using 2nd order Lagrangian perturbation theory, \texttt{2LPT} \citep{Scoccimarro1998, Crocce+2006} and the cosmological parameters are taken from Planck constraint
\citep{Planck2016}: ($h$, $\Omega_{\rm b0}$, $\Omega_{\rm c0}$, $\Omega_{\Lambda 0}$, $n_{\rm s}$, $\sigma_8$) = (0.6727, 0.0493, 0.2653, 0.6844, 0.9645, 0.831).
The box size of the simulation is 
$L_{\text{box}} = 2000$ \mpc\ with $2048^3$ particles in comoving periodic cubes.
They provide snapshots at 21 different redshifts at $0 < z < 1.48$; for each, there are 14 independent different realisations. We use the snapshots at $z=0.251, 0.617$, and 1.03.
To identify dark matter halos, 
the ROCKSTAR \citep{Behroozi+2013} finder was used, in addition to \texttt{Subfind} \citep{Springel+2001} to study the dependence of the halo statistics on the finder. 

Then, we employ the Halo Occupation Distribution (HOD) to 
populate galaxies within a halo.
Following 
\cite{ishikawa+2021}, we generate the mock catalogue assuming the Luminous Red Galaxies (LRGs) \citep{Oguri+2018} observed using the Subaru Hyper Suprime-Cam (HSC) \citep{Aihara+2018}.
The total number of LRGs within a dark matter halo of
mass $M_\text{h}$ can be decomposed into the number of central ($N_{\text{cen}}$) and satellites ($N_{\text{sat}}$) \citep{Zheng+2005}, 
\begin{align}
    N_{\text{tot}} (M_\text{h}) &= N_{\text{cen}}(M_\text{h}) [1 + N_{\text{sat}}(M_\text{h})], \\
    N_{\text{cen}}(M_\text{h}) &= \dfrac{1}{2} \left[1 + \text{erf} \left(\dfrac{\log_{10}M_\text{h} - \log_{10}M_{\text{min}}}{\sigma_{\log M}}\right)\right], \\
    N_{\text{sat}}(M_\text{h}) &= \left(\dfrac{M_\text{h} - \kappa M_{\text{min}}}{M_1}\right)^{\alpha_{\text{HOD}}},
\end{align}
where $M_{\text{min}}$, $M_1$, $\kappa$, $\alpha_{\text{HOD}}$, and $\sigma_{\log M}$ are HOD parameters.
For the values of HOD parameters, we follow \cite{ishikawa+2021} corresponding to the stellar mass limits of $10^{11} M_{\sun}/h^2$ and  $10^{10.25} M_{\sun}/h^2$.

Once the galaxies are populated in the box, we divide the simulation box into 8 regions where the box size of the subregion is 1000 \mpc\, on a side. This allows us to increase the number of quasi independent realisations from 14 to 112, which will avoid the noisy measurement of the covariance matrix \citep{Hartlap+2007}.

\subsection{Effect of photometric redshift}
\label{ssec:photoz}
In this section, we describe the way to incorporate the photo-$z$ uncertainty into the mock data.
We first convert the real-space galaxy positions to those in redshift space,
\begin{align}
    s(z) = r(z) + \dfrac{v_{\text{z}}(1+z)}{H(z)}, \label{eq:RSD}
\end{align}
where $v_{\text{z}}$ is the peculiar velocity of the galaxy in the LOS direction. Hereafter, $s(z)$ is a ground truth of the galaxy radial position in redshift space, and we add a fluctuation around it according to the photometric redshift uncertainty.
We refer to the comoving radial distance uncertainty due to the photo-$z$ uncertainty as $\Delta r$ and can be computed as
\begin{align}
    \Delta r(z) = \dfrac{1}{2}\int_{z-\sigma_{\text{p}}}^{z+\sigma_{\text{p}}} \dfrac{c \mathrm{d} z}{H(z)}, 
    \label{eq:Deltacomz}
\end{align}
where $\sigma_{\text{p}} = \sigma_z(1+z)$ is the redshift uncertainty in photometric observation.
For example, the 1\% photo-$z$ error at $z=0.251$ reads photo-$z$ uncertainty of $\sigma_{\rm P}=0.0125$ and thus $\Delta r=32.9$ \mpc. The full correspondence 
can be found in Table \ref{tab:photoz_error}. The position of the galaxy is randomly placed based on the probability of ${\cal N}(\hat{r}_z, \Delta r)$, where $\hat{r}_z$ is the z-axis coordinate of the simulation in the redshift space (Eq.~\ref{eq:RSD}).
\begin{table}
  \begin{center}
      \caption{
      One sigma radial distance uncertainty due to the photo-$z$ accuracy in units of \mpc.}
      \label{tab:photoz_error}
      \begin{tabular}{c c c c} \hline
         redshift $z$ & \multicolumn{3}{c}{photo-$z$ error} \\ \hline 
         & 1\% & 2\% & 3\% \\ \hline \hline
        0.251 & 32.9 & 65.7  & 98.6 \\
        0.617 & 34.1 & 68.2  & 102.5 \\
        1.03  & 33.3 & 66.8  & 100.2 \\
        \hline
      \end{tabular}
  \end{center}
\end{table}
As can be seen
in Table \ref{tab:photoz_error}, the photo-$z$ 3\% error in comoving
distance is comparable to the BAO scale, and thus, we limit our analysis for the photo-$z$ uncertainties better than 3\%.

\subsection{Mock measurement}
\label{ssec:measurement}

Here, we describe the correlation function measurement from the mock simulations. We measure the correlation function in redshift space, but we do not distinguish the direction of the separation since our theoretical model is integrated over the angular direction. We employ the Landy Szaly estimator \citep{Landy_Szalay1993} to measure the correlation function within the sub-box of the simulation.
We generate random points in the simulation cubic box with the number of random points 10 times larger than that of the galaxies. We average correlation functions measured in 112 simulation sub-boxes of 1000 \mpc,
\begin{align}
    \ave{\xi_{\rm mock}(r)}
    =
    \frac{1}{N_{\rm mock}}\sum_i \xi_{\rm mock}^i(r)
\end{align}
The covariance can be measured over 112 mock simulations,
\begin{align}
    \text{Cov}(r_1, r_2) &=& \dfrac{1}{N_{\text{mock}} - 1}\sum_i 
    \left[ \xi^i_{\rm mock}(r_1) - \ave{\xi_{\text{mock}}(r_1)} \right] \nonumber\\
    &&\times\left[ \xi^i_{\rm mock}(r_2) - \ave{\xi_{\text{mock}}(r_2)} \right], \label{eq:covariance}
\end{align}
where $N_{\text{mock}}$ is the number of mock realisations. 
The covariance corresponds to a survey volume of 1~[\gpc\,]$^3$, and we will simply scale this by $V_{\rm survey}$ to match to a hypothetical survey setting. For example, the covariance matrix for Vera Rubin/LSST survey at $z\sim 1$ provides $V_{\rm LSST}\sim 11~[$\gpc$]^3$ survey volume and thus the expected covariance will be scaled as
\begin{align}
    {\rm Cov_{\rm LSST}} = \frac{1 \,[h^{-1}\text{Gpc}]^3}{V_{\rm LSST}} {\rm Cov}. \label{eq:scaled_cov}
\end{align}
In section \ref{ssec:LSST-like_covariance}, we will discuss the expected constraint on the $\alpha$ parameter using the rescaled covariance assuming an LSST-like survey.

\section{Results}
\label{sec:results}
In this section, we show the fitting results and constraints on $\alpha$.
As described in \S~\ref{ssec:model}, we fit the model to the measured correlation function. The measured correlation function is the average of the correlation functions measured from $14\times 8$ mock realisations, and we compute the covariance matrix using the same samples.
In each subsection, we change the assumption about photometric redshift and the set-up of the templates. 
First, we assume that we have an accurate estimate of photometric redshift uncertainty for the galaxies with $M_* \ge 10^{11} M_{\sun} / h^2$ at $z=0.251$, $0.617$, and $1.03$.
Additionally, we investigate whether the number density of galaxies matters for the covariance matrix, which is crucial for the accuracy of the measured $\alpha$ value. To test this, we changed the selection of galaxies from $M_* \ge 10^{11} M_{\sun}/h^2$ to $M_* \ge 10^{10.25} M_{\sun} / h^2$.
Then, we explore whether assuming a wrong size of uncertainty on photometric redshift can bias the constraint on $\alpha$.
At last, we investigate whether cross-correlation of the photometric galaxy sample with the spectroscopic galaxy can improve the precision and accuracy on the constraint of $\alpha$.

\subsection{Correlation function and constraining BAO parameter}
\label{ssec:bao_measure_perror_isknown}
\begin{figure*}
    \begin{center}
        \begin{tabular}{ccc}
            \includegraphics[width=0.31\linewidth,keepaspectratio]{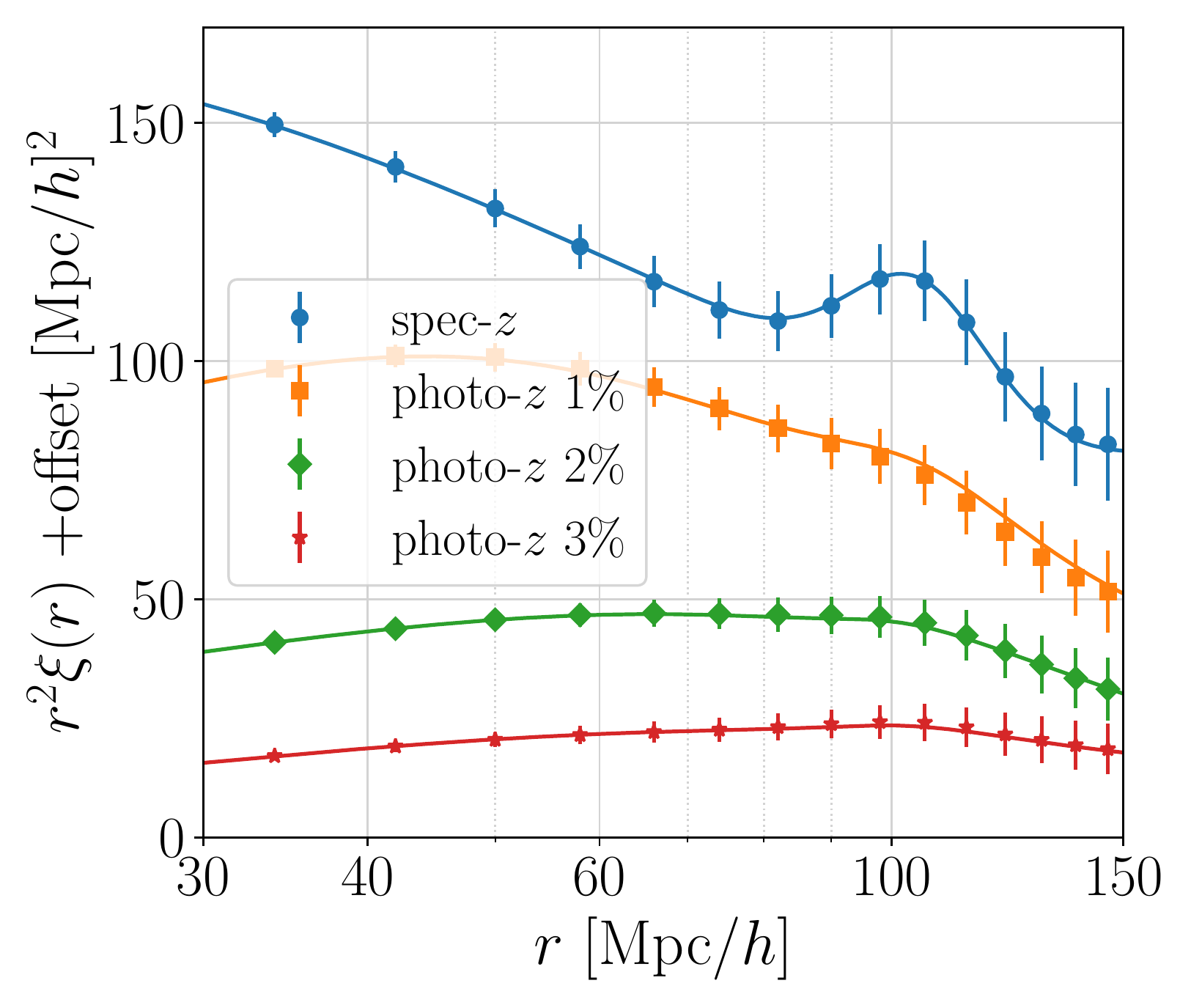}&
            \includegraphics[width=0.31\linewidth,keepaspectratio]{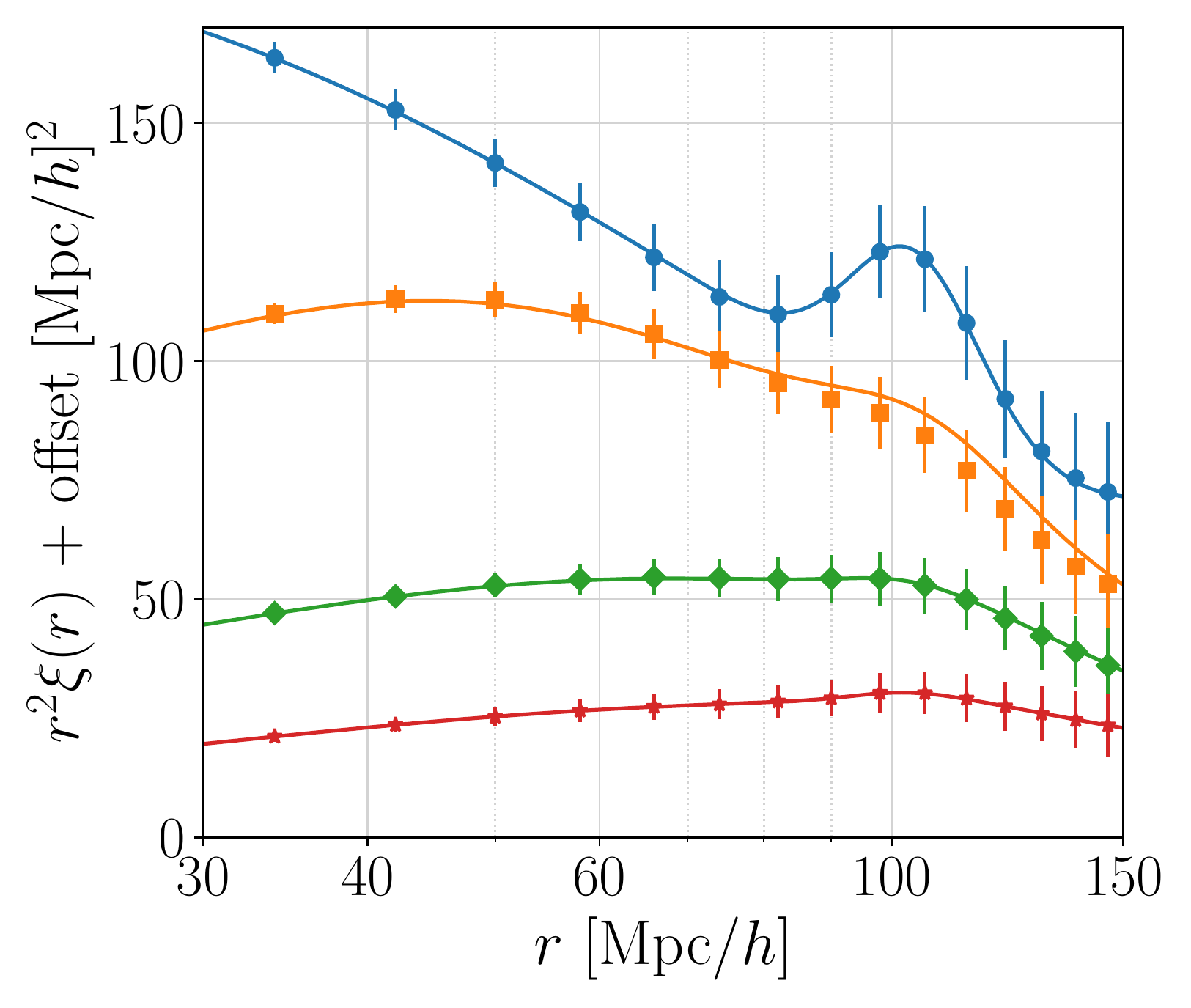}&
            \includegraphics[width=0.31\linewidth,keepaspectratio]{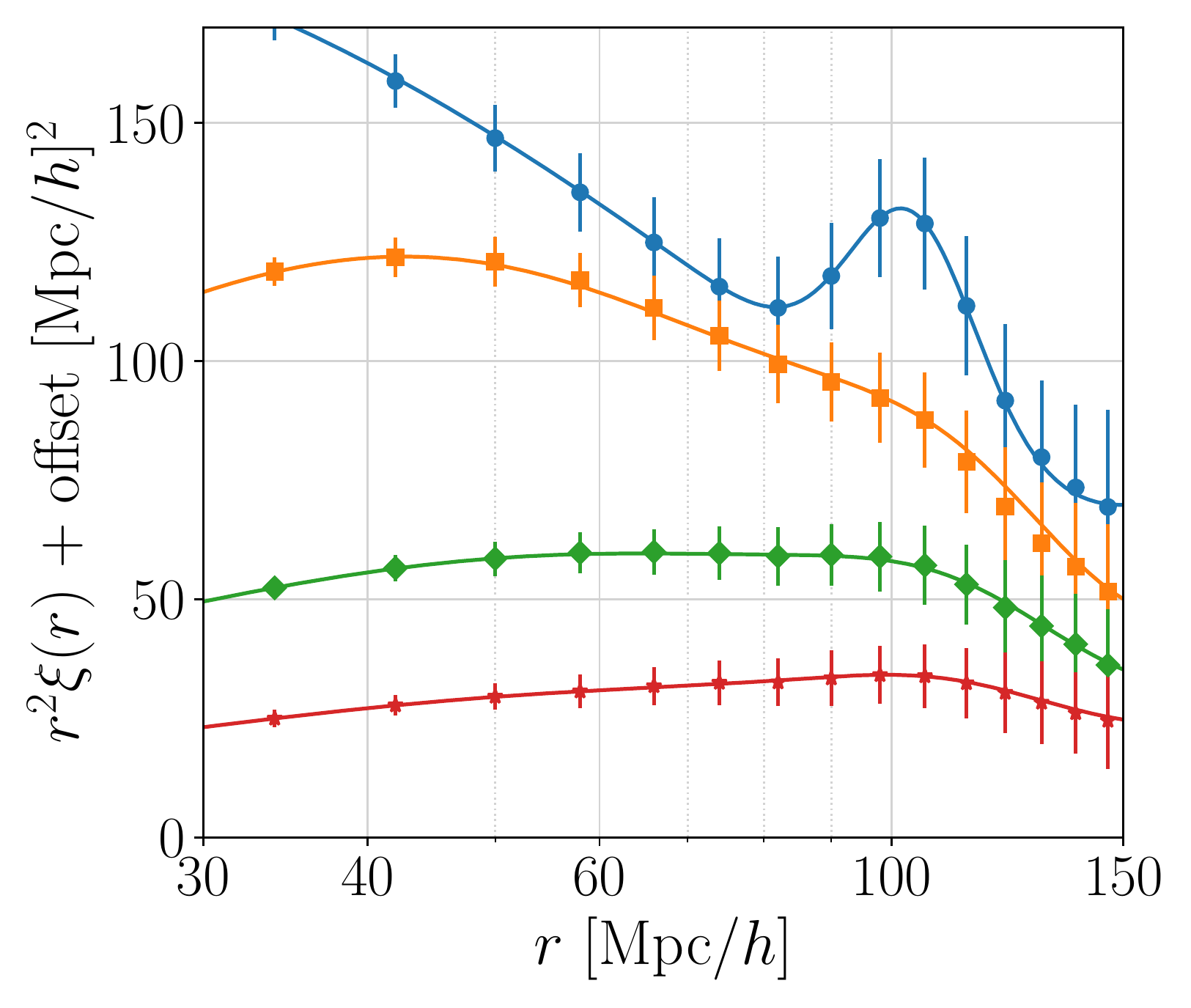}\\
            (a) $z=0.251$ &
            (b) $z=0.617$ &
            (c) $z=1.03$ \\
        \end{tabular}
        \caption{
        redshift space correlation functions as a function of photometric redshift uncertainties at three different redshifts ($z=0.251, 0.617$, and $1.03$). Points with error bars are the measured correlation functions from 112 mock realisations, and the solid lines are the best-fit models using the template with a correctly assumed photometric redshift uncertainty.
        Galaxies are selected with $M_* \ge 10^{11} M_{\sun} / h^2$ using the result of \citep{ishikawa+2021}. The amplitude of the correlation functions is shifted for illustrative purposes. 
        }
        \label{fig:eachfit_result_SML11}
    \end{center}
\end{figure*}
In most photometric galaxy surveys, the calibration data accurately estimate the photometric redshift uncertainty. Therefore, we first assume that we understand the size of the photometric redshift uncertainty accurately. More specifically, we fix the photometric redshift uncertainties $\sigma_P$ to the correct value in evaluating Eq.~\eqref{eq:xi_int}. We use the mock galaxies with $M_* > 10^{11}M_{\sun}/h^2$ at redshifts $z=0.251$, $0.617$, and $1.03$.
Fig.~\ref{fig:eachfit_result_SML11} shows the redshift space correlation functions at $z=0.251, 0.617$, and $1.03$ with various sizes of photometric redshift uncertainties. Compared to the correlation functions measured from spectroscopic galaxies, the BAO peak is smeared out even in the case of photometric redshift uncertainty of $\sigma_P=1\%$. However, our model template, which includes the photometric redshift uncertainty, can correctly fit the measured correlation function, and they all agree well within $1\sigma$.
As shown in the figures, the size of errors increases at higher redshift since the number density gets smaller due to the fixed stellar mass limit in  \S~\ref{ssec:bao_measure_perror_isknown}. Another feature is that the error size is anti-correlated with the size of the photometric redshift uncertainty. 
\begin{figure}
    \begin{center}
        \includegraphics[width=\linewidth,keepaspectratio]{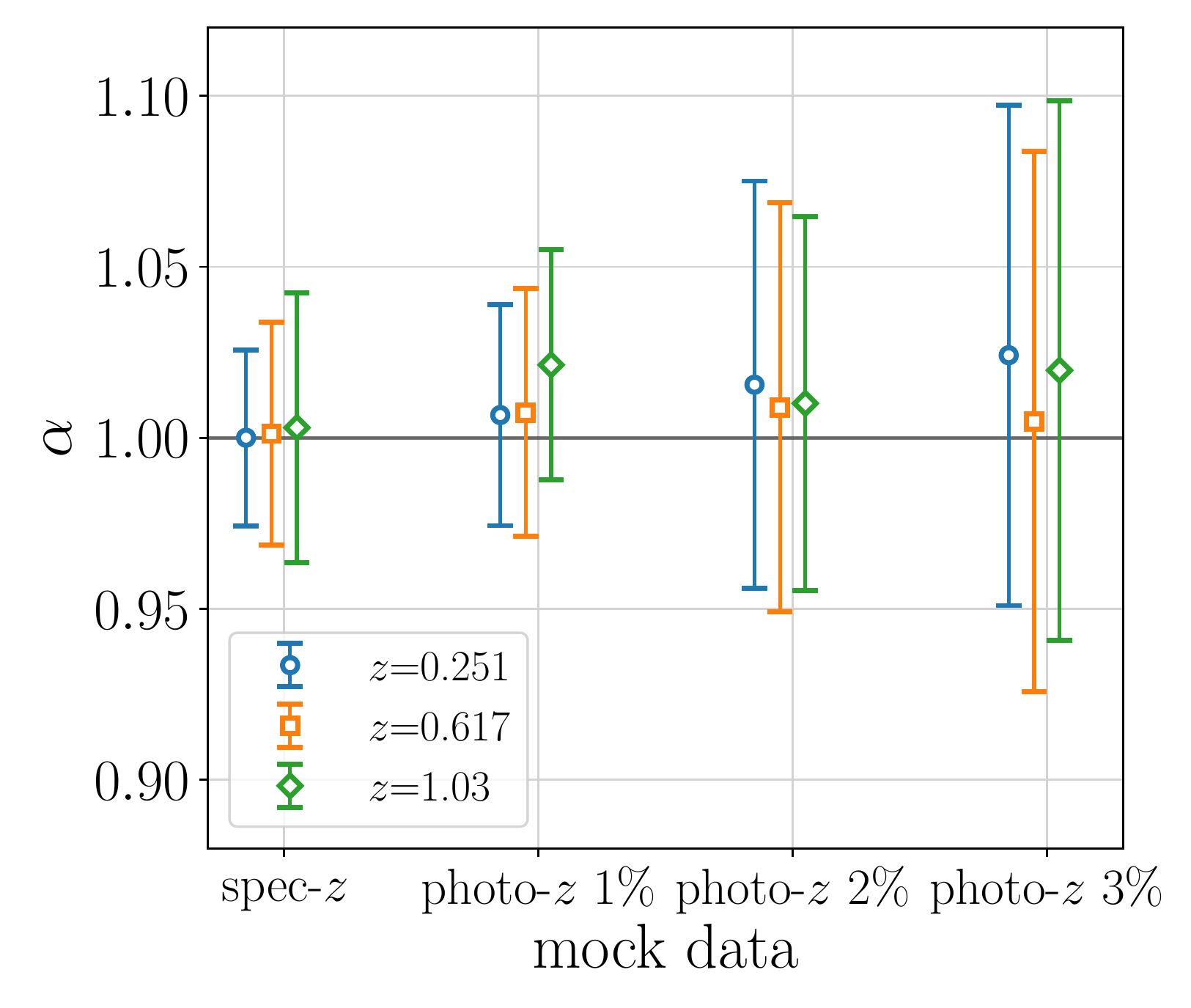}
        \caption{
        Constraints on $\alpha$ parameter for $M_*>10^{11}M_{\sun}/h^2$ sample as a function of the photo-$z$ uncertainty at three different redshifts, $z=0.251$(blue-circles), $z=0.617$ (orange-squares), and $z=1.03$ (green-diamonds), respectively.
        }
        \label{fig:alpha_value_SML11_each_fit}
    \end{center}
\end{figure}

Fig.~\ref{fig:alpha_value_SML11_each_fit} shows the best-fit values of $\alpha$ and their confidence intervals marginalised over all other nuisance parameters. The figure shows that the constraining power on $\alpha$ decreases as photo-$z$ uncertainty increases, which is expected as the BAO peak is almost undetectable with large photo-$z$ uncertainty. For almost all the cases, the shifts of the BAO peak $\alpha$ are accurately constrained. In the case of photo-$z$ uncertainty of $1\%$ at $z=1.03$, the best-fit value of $\alpha$ is slightly biased, though it is still within $1\sigma$, and we suspect this is due to a low number density of galaxies and a noisy covariance matrix measured from these samples.
     
\begin{figure}
   \begin{center}
       \includegraphics[width=\linewidth,keepaspectratio]{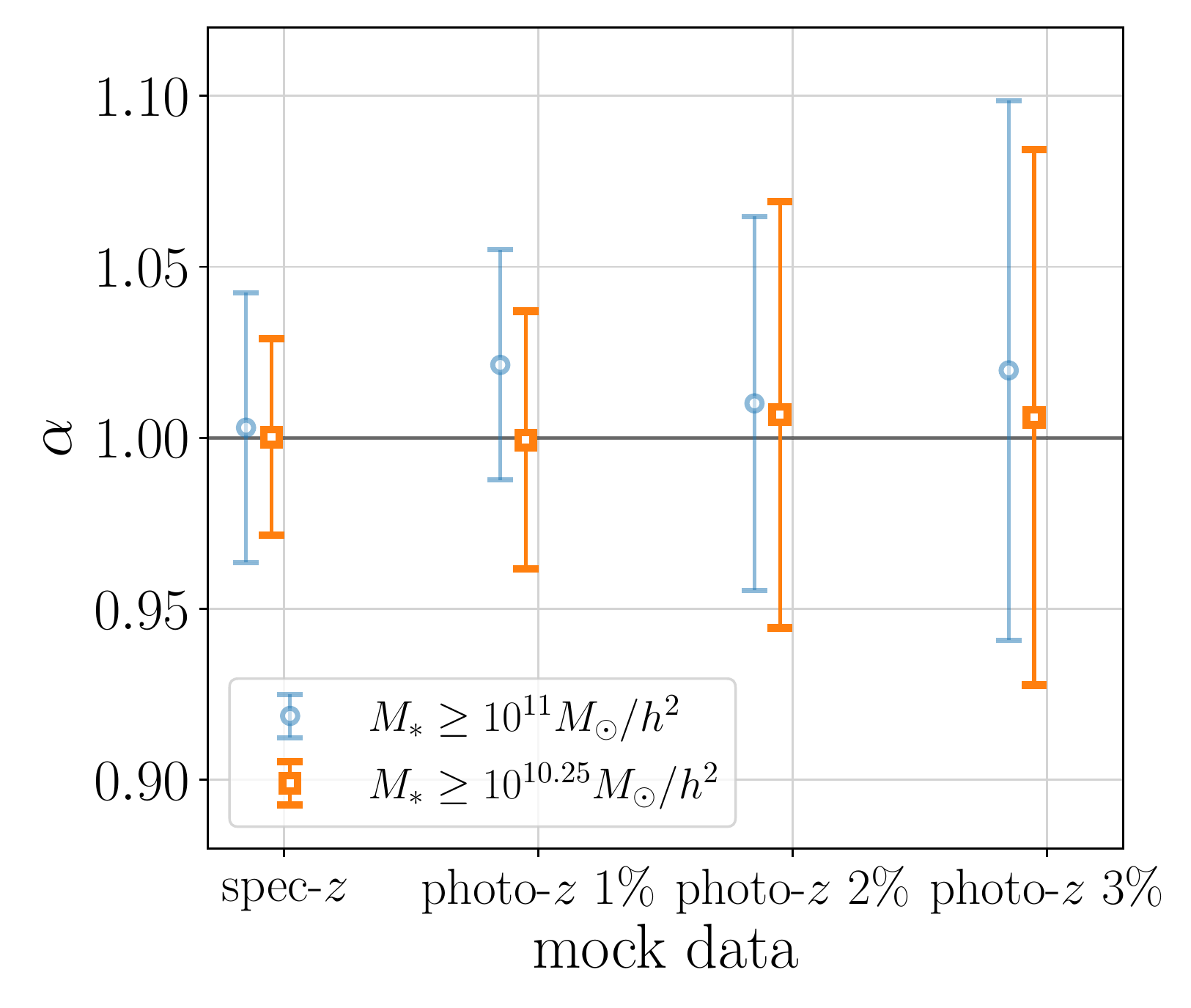}
       \caption{
       The best-fit values of $\alpha$ at $z=1.03$ for different stellar mass limit samples. The stellar mass limit $M_* \ge 10^{11} M_{\sun} / h^2$ is in blue and the one $M_* \ge 10^{10.25} M_{\sun} / h^2$ is in orange. These error bars represent 95\% confidence intervals.}
       \label{fig:alpha_value_S004_compared_SML11}
   \end{center}
\end{figure}

To test this, we decide to use a sample of galaxies with $M_* \geq 10^{10.25}M_{\sun}/h^2$ at $z=1.03$. 
We again follow \cite{ishikawa+2021} to generate mock samples with HOD parameters for the galaxies with $M_* \geq 10^{10.25}M_{\sun}/h^2$. Compared with the galaxy sample with $M_* \geq 10^{11} M_{\sun}/h^2$, the number density of the sample increases from $2 \times 10^{-4}$ to $2 \times 10^{-3}$[\mpc]$^{-3}$.    
Fig.~\ref{fig:alpha_value_S004_compared_SML11} shows the constraints on $\alpha$ for the case of galaxies with $M_* \geq 10^{10.25}M_{\sun}/h^2$. Now, the best-fit value of $\alpha$ for the case of the photo-$z$ error 1\% is not biased. 
This change in the constraint on $\alpha$ may be due to a lower galaxy bias or a covariance matrix with reduced noise.
A lower galaxy bias can possibly give an unbiased constraint on $\alpha$ because the shift of the BAO peak depends on galaxy bias \citep{PadmanabhanWhite_2009,Seo_2005,Mehta_2011} and a higher galaxy bias shifts the BAO peak more. 
Another possibility is that a covariance matrix measured from galaxies with $M_* \geq 10^{11}M_{\sun} / h^2$ is noisy due to its low number density.
To test which one explains the change in the result shown in Fig.~\ref{fig:alpha_value_S004_compared_SML11}, we use the measurement from galaxies with $M_* \geq 10^{11} M_{\sun}/h^2$ but constrain the value of $\alpha$ using the covariance matrix measured from galaxies with $M_* \geq 10^{10.25} M_{\sun}/h^2$. 
This test gives an unbiased value of $\alpha$ implying that the slightly biased $\alpha$ value shown in Fig.~\ref{fig:alpha_value_SML11_each_fit} is due to a noisy covariance matrix, not because of a larger galaxy bias. 
Another thing to note from Fig.~\ref{fig:alpha_value_S004_compared_SML11} is that the error becomes slightly larger for the case of photo-$z$ error 2\% and 3\% by using a covariance matrix measured by a galaxy sample with a higher number density.
It implies that a noisy covariance gives an underestimated error for the constraint of $\alpha$.
    
\begin{figure*}
    \begin{center}
        \includegraphics[width=\linewidth,keepaspectratio]{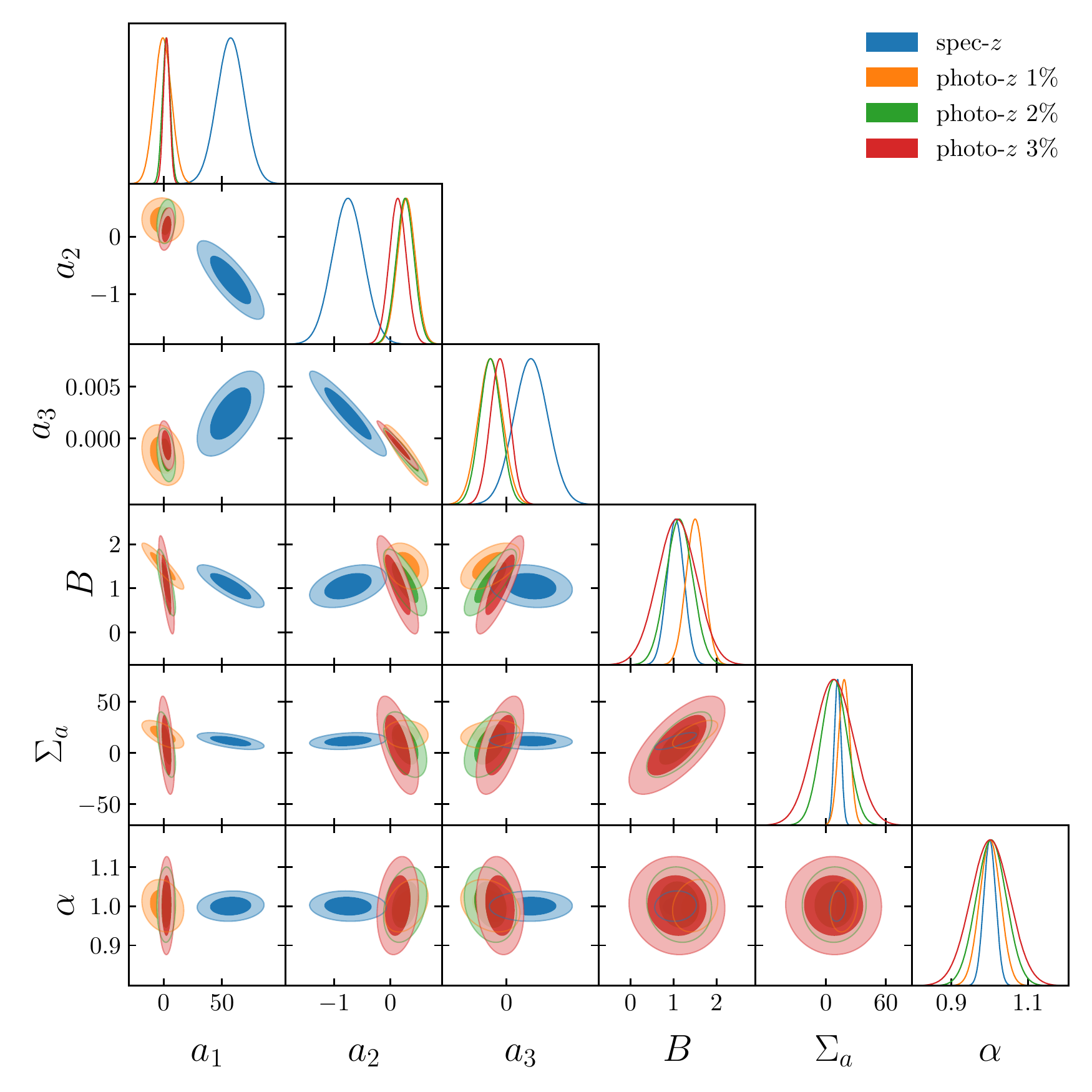}
        \caption{A triangle plot of the full parameter constraints for the samples at $z=1.03$ fixed by the stellar mass limit being $M_* \ge 10^{10.25} M_{\sun}/h^2$ with being overwritten by different redshift accuracies. As mentioned above, the spec-$z$ data is in blue, the photo-$z$ 1\% data is in orange, the photo-$z$ 2\% is in green, and the 3\% is in red.}
        \label{fig:triangle_plot_SML1025_known_photoz_overwrite}
    \end{center}
\end{figure*}

Next, we study how the $\alpha$ parameter degenerates with other nuisance parameters to explore how the degeneracy between parameters depends on the size of photo-$z$ accuracy.
Fig.~\ref{fig:triangle_plot_SML1025_known_photoz_overwrite} shows the full parameter constraints for the samples at $z=1.03$ with different photo-$z$ uncertainties. 
As the figure shows, the constraints on nuisance parameters are different for spectroscopic and photometric galaxies. However, these constraints from photometric galaxies do not show a strong dependence on the size of the uncertainty. For the case of spectroscopic galaxies, the best-fit values of the nuisance parameters $a_1$, $a_2$, and $a_3$ are non-zero, while the best-fit values of these parameters are almost zero for the case of galaxies with photometric redshift uncertainties. 
These parameters are meant to model the scale dependence of galaxy bias on small scales due to non-linear evolution. 
Therefore, it makes sense that the best-fit values for these parameters are almost zero for the case of photometric galaxies since photometric redshift uncertainties smear out the information on small-scale clustering.
However, despite the fact that the best-fit values of the nuisance parameters are almost zero, we find that discarding these parameters in our model biases the constraint on $\alpha$, and it is still crucial to have them to obtain an unbiased measurement of $\alpha$.
Another notable feature of this figure is that the degeneracy between $\alpha$ and other nuisance parameters is weak for both spectroscopic and photometric galaxies. However, the direction of the contour is opposite. 
While $\alpha$ is constrained well and other nuisance parameters have a relatively broad scatter for the case of spectroscopic galaxies, the trend is the opposite for the case of photometric galaxies.
This may be because our model is based on the template function primarily for spectroscopic galaxies. There might be a better way to parameterise the template function for photometric galaxies.
However, in both cases, the best-fit value of $\alpha$ always stays around $\alpha$ = 1, meaning that the model is quite stable and can robustly constrain $\alpha$.
  
This result from the galaxy sample $M_* \ge 10^{10.25} M_{\sun}/h^2$ at $z=1.03$ is comparable to \cite{Chan+2022b} when box size and the number of realisations are scaled accordingly.
Compared with ours, \cite{Chan+2022b} has a similar halo mass limit, but their box size is 1.5 times larger on a side (i.e. their simulation box volumes 1.5$^3$ [\gpc$^3$]).
Therefore, the scaled error bar for a 3\% photo-$z$ (detailed in Table~\ref{tab:fitting_results_in_Sec_Results}) is 0.023 at 68\% confidence level.
Another factor that would cause difference with our analysis is
the number of realisations used for the estimation of the covariance.
Figure 2 of \cite{Chan+2022b} implies that it may seem difficult to measure the BAO peak scale using a 3D correlation function because the BAO scale is strongly dependent on $\mu$. However, if we understand the amount of photo-$z$ uncertainty and it is correctly incorporated in the model, the BAO peak, integrated over the $\mu$ direction reproduces the correct BAO scale at high accuracy. Therefore, although the projected correlation function is widely used for the photo-$z$ BAO, the 3D correlation function offers a complementary estimation of the BAO and a comparable ability to constrain the BAO scale.

\subsection{Unknown photo-$z$ accuracy}
\label{ssec:bao_measure_perror_isunknown}
In this section, we further explore the case when we do not understand the photo-$z$ accuracy beforehand, or we misestimate it. The most straightforward way to test this is to make the photo-$z$ accuracy parameter $\sigma_P$ free. However, the parameter $\sigma_P$ is highly degenerate with the smoothing parameter $\Sigma_a$; therefore, the fitting does not work properly. Here instead, we fix the photo-$z$ parameter $\sigma_P$, but set it to the wrong value and quantify how assuming a wrong photo-$z$ accuracy impacts the constraint on $\alpha$.
    
\begin{figure}
    \begin{center}
        \includegraphics[width=\linewidth,keepaspectratio]{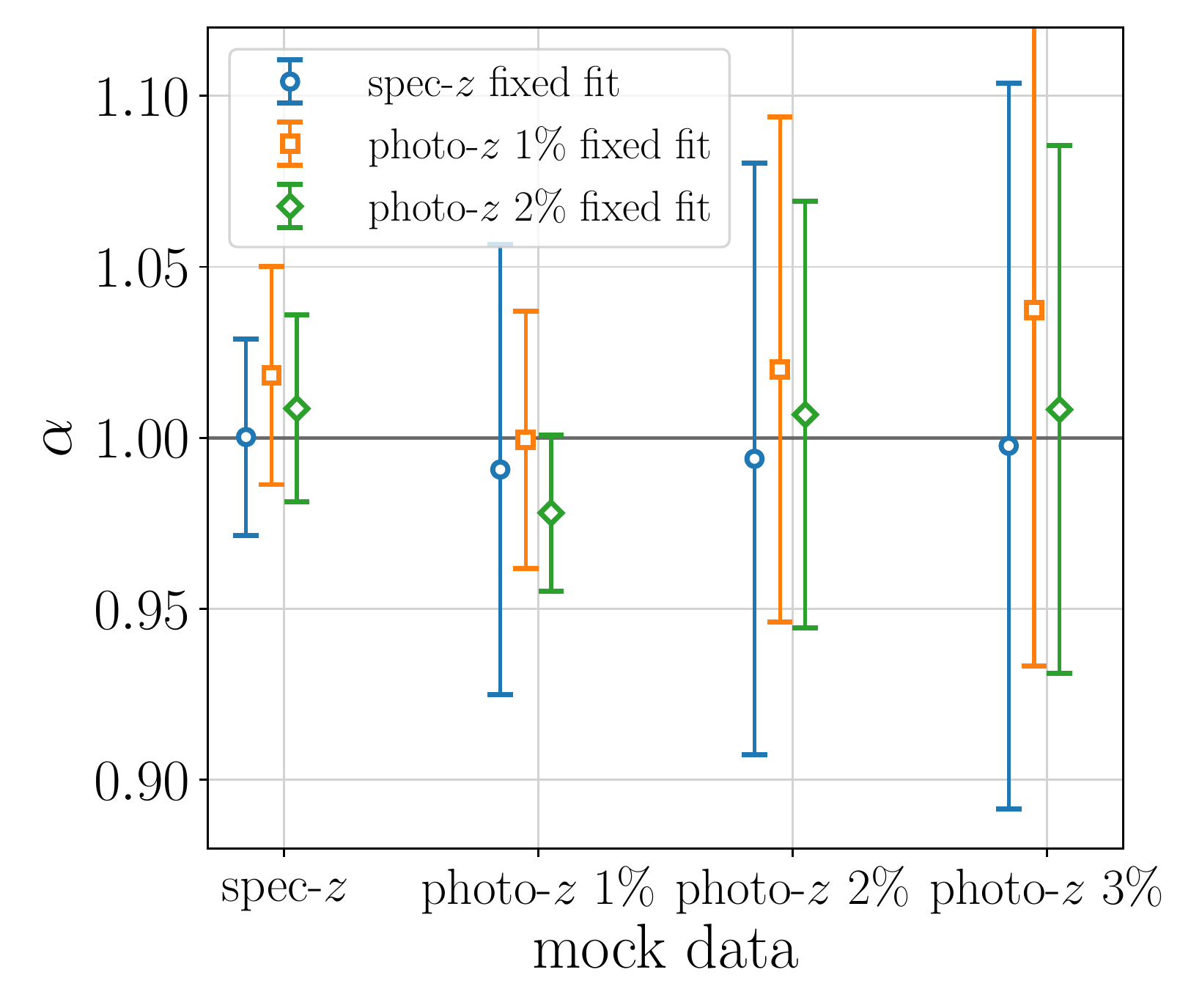} \caption{The best-fit values of $\alpha$ at $z=1.03$ fixed by the stellar mass limit being $M_* \ge 10^{10.25} M_{\sun} / h^2$ with different fixed photo-$z$ accuracy parameters. The data in the case we fix spec-$z$ template is in blue, the data in the case we fix photo-$z$ 1\% template is in orange, and the case with photo-$z$ 2\% template is in green. These error bars represent 95\% confidence intervals.}
        \label{fig:alpha_value_S004_fixed_photoz_fit}
    \end{center}
\end{figure}
Fig.~\ref{fig:alpha_value_S004_fixed_photoz_fit} shows the constraints on $\alpha$ for the galaxy sample at $z=1.03$ as a function of photo-$z$ uncertainty. Different colour corresponds to a different assumption of photo-$z$ accuracy used in the fitting model.
Let us first focus on the results fitted by the model assuming spectroscopic galaxies. The constraining power on $\alpha$ becomes weaker as the induced photo-$z$ error increases. The reason is due to the fact the photometric redshift uncertainty smears out the BAO peak while the fitting model still has a sharp BAO peak. It is worth noting that the constraints on $\alpha$ are still unbiased.
Then we compare those results for the case of using the model assuming the photo-$z$ uncertainty of 1\% and 2\%. When the photometric redshift uncertainty is underestimated, the error gets larger compared to the case with a consistent model template. On the other hand, if the photo-$z$ uncertainty is overestimated, the constraint on $\alpha$ can be biased.

\subsection{Cross-correlation between the spec-$z$ catalogue and the photo-$z$ catalogue}
\label{ssec:bao_measure_perror_isknown_cross}
An alternative way to harness the photometric sample is to take a cross-correlation function with the spec-$z$ galaxy sample \citep{Nishizawa+2013, Patej+2018, Zarrouk+2021}.
We assume the sparse spec-$z$ galaxy sample of $M_* \ge 10^{11}M_{\sun}/h^2$ with no photo-$z$ uncertainty, and the photometric galaxy sample of $M_* \ge 10^{10.25}M_{\sun}/h^2$ sample with photo-$z$ uncertainties.
\begin{figure}
    \begin{center}
        \includegraphics[width=\linewidth,keepaspectratio]{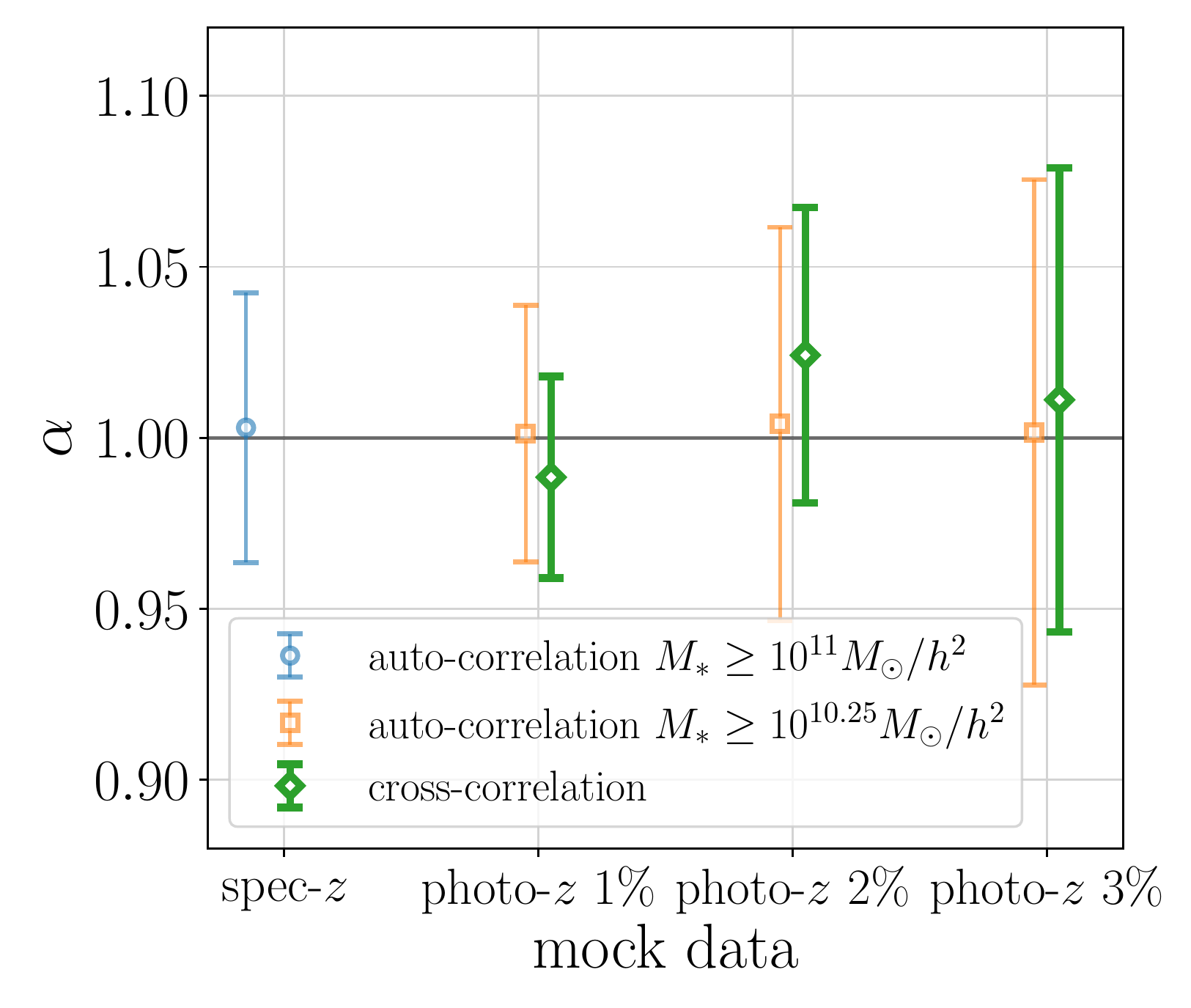}
        \caption{The best-fit values of $\alpha$ at $z$=1.03 using cross-correlation (green-diamond). The same ones in blue and orange are in Fig.~\ref{fig:alpha_value_S004_compared_SML11} and these data in blue and orange are thinly plotted for comparison. The cross-correlation in green consists of the spec-$z$ catalogue of the stellar mass limit $M_* \ge 10^{11} M_{\sun}/h^2$ and the photo-$z$ catalogue of the stellar mass limit $M_* \ge 10^{10.25} M_{\sun}/h^2$. The error bars are with 95\% confidence (2$\sigma$). }
        \label{fig:alpha_value_S004_cross}
    \end{center}
\end{figure}
Fig.~\ref{fig:alpha_value_S004_cross} shows the constraints on $\alpha$ using auto-correlation functions of the spec-$z$ galaxy sample, the photo-$z$ galaxy sample, and cross-correlation functions between the spec-$z$ and photo-$z$ galaxy samples.
The constraint on $\alpha$ using the cross-correlation function is expected to be better since the spec-$z$ galaxy sample can calibrate the photo-$z$ uncertainty of the photo-$z$ galaxy sample.
For example, the error of the $\alpha$ parameter in photo-$z$ 1\% cross-correlation function (green) is 32.3\% smaller than that of photo-$z$ 1\% auto-correlation function.
As mentioned in the previous section, the bias of the best-fit $\alpha$ parameter in cross-correlation functions is caused by the noise in the covariance matrix measured from the sparse spectroscopic galaxy sample. This is the same reason for the biased $\alpha$ value measured from the photo-$z$ 1\% galaxy sample at $z=1.03$ as shown in Fig.~\ref{fig:alpha_value_SML11_each_fit}.

We summarise all the results presented in this section in Table~\ref{tab:fitting_results_in_Sec_Results}.
\begin{table*}
  \begin{center}
      \caption{The fitting results in \S~\ref{sec:results}. 
      These rows correspond to each Section and $z=1.03$ ($M_* \ge 10^{10.25} M_{\sun}/h^2$) in \S~\ref{ssec:bao_measure_perror_isknown} is fiducial set-up in this paper.
      The cross-correlation is conducted between spec-$z$ and each photo-$z$ data, so the column of spec-$z$ data is blank.
      These columns correspond to the data which is used, and the column of spec-$z$ data is for comparison.
      These values are the best fits using least $\chi^2$ fitting, and these errors are marginalised over the covariance of fitting parameters space generated after fitting.
      The error bars are with 95\% confidence.
      }
      \label{tab:fitting_results_in_Sec_Results}
      \begin{tabular}{c c c c c c c} \hline
        & stellar mass limit & spec-$z$ data & \multicolumn{3}{c}{photo-$z$ data} & ref. \\ \hline 
        & $M_* \ge 10^x M_{\sun}/h^2$ & & 1\% error & 2\% error & 3\% error & \\ \hline 
        $z=0.251$ & $x=11.00$ 
        & 1.000 $\pm$ 0.026 & 1.007 $\pm$ 0.032 & 1.016 $\pm$ 0.060 & 1.024 $\pm$ 0.073 & 
        \multirow{3}{*}{\S~\ref{ssec:bao_measure_perror_isknown}} \\
        $z=0.617$ & $x=11.00$
        & 1.001 $\pm$ 0.033 & 1.007 $\pm$ 0.036 & 1.009 $\pm$ 0.060 & 1.005 $\pm$ 0.079 & \\
        $z=1.03$  & $x=11.00$
        & 1.003 $\pm$ 0.039 & 1.021 $\pm$ 0.034 & 1.010 $\pm$ 0.055 & 1.020 $\pm$ 0.079 & \\ \hline \hline
        $z=1.03$  & $x=10.25$ 
        & 1.000 $\pm$ 0.029 & 0.999 $\pm$ 0.038 & 1.007 $\pm$ 0.062 & 1.006 $\pm$ 0.078 & 
        \S~\ref{ssec:bao_measure_perror_isknown}  \\ \hline \hline
        spec-$z$ fix template & $x=10.25$ 
        & 1.000 $\pm$ 0.029 & 0.991 $\pm$ 0.066 & 0.994 $\pm$ 0.086 & 0.998 $\pm$ 0.106 & 
        \multirow{3}{*}{\S~\ref{ssec:bao_measure_perror_isunknown}} \\
        photo-$z$ 1\% fix template & $x=10.25$ 
        & 1.018 $\pm$ 0.032 & 0.999 $\pm$ 0.038 & 1.020 $\pm$ 0.074 & 1.037 $\pm$ 0.104 & \\
        photo-$z$ 2\% fix template & $x=10.25$ 
        & 1.009 $\pm$ 0.027 & 0.978 $\pm$ 0.023 & 1.007 $\pm$ 0.062 & 1.008 $\pm$ 0.077 & \\ \hline
        \multirow{2}{*}{cross-correlation with spec-$z$} & $x=10.25$ (photo-$z$)
        & & \multirow{2}{*}{0.990 $\pm$ 0.025} & \multirow{2}{*}{1.027 $\pm$ 0.047} & \multirow{2}{*}{1.009 $\pm$ 0.077} & \multirow{2}{*}{\S~\ref{ssec:bao_measure_perror_isknown_cross}} \\
        & $x=11.00$  (spec-$z$) & \multicolumn{5}{c}{} \\ \hline
      \end{tabular}
  \end{center}
\end{table*}

\section{Discussion}
\label{sec:discussion}

\subsection{Effect of incorrect cosmological models}
\label{ssec:incorrect_cosmology}
\begin{figure*}
    \begin{center}
        \begin{tabular}{ccc}
            \includegraphics[width=0.31\linewidth,keepaspectratio]{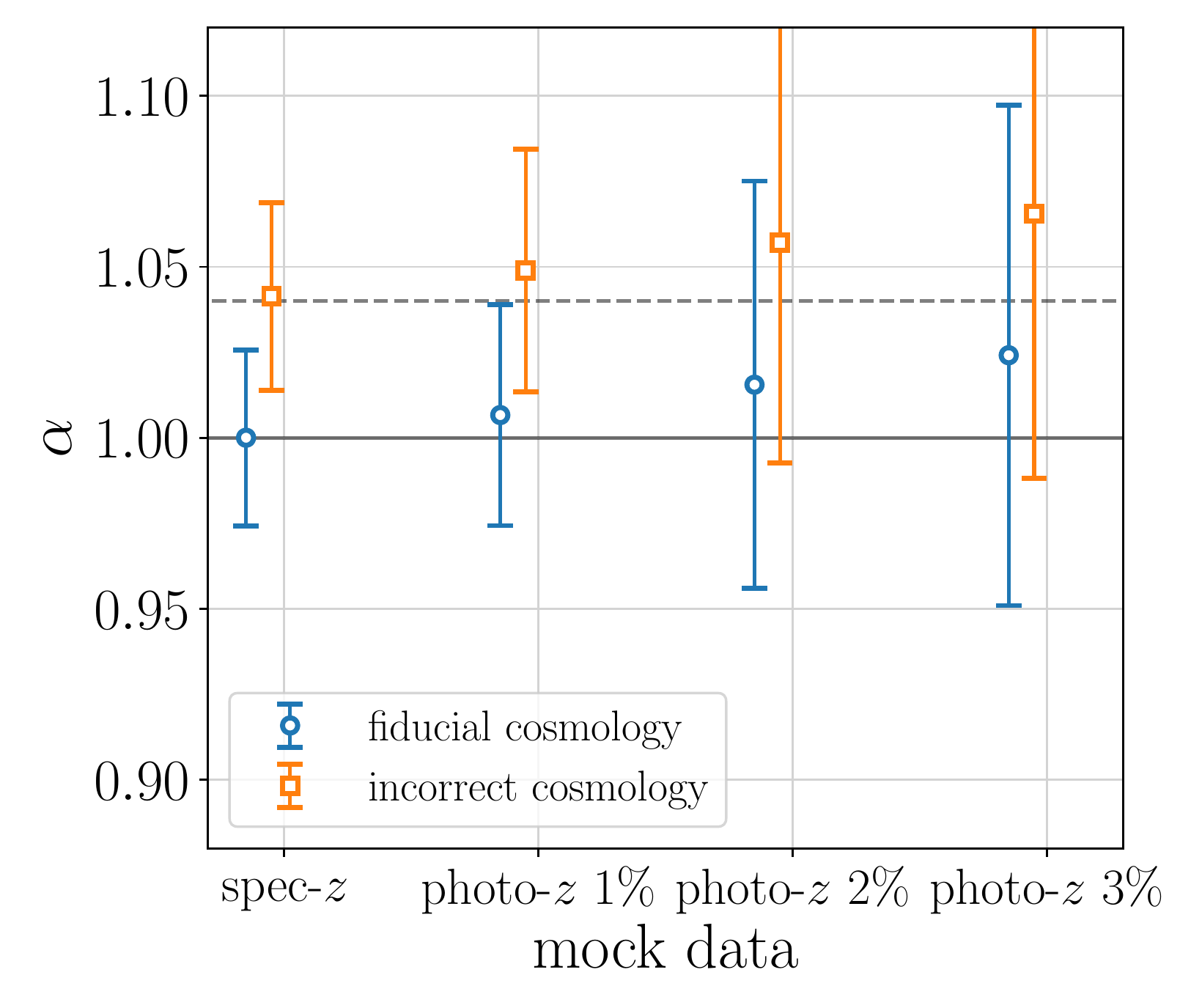}&
            \includegraphics[width=0.31\linewidth,keepaspectratio]{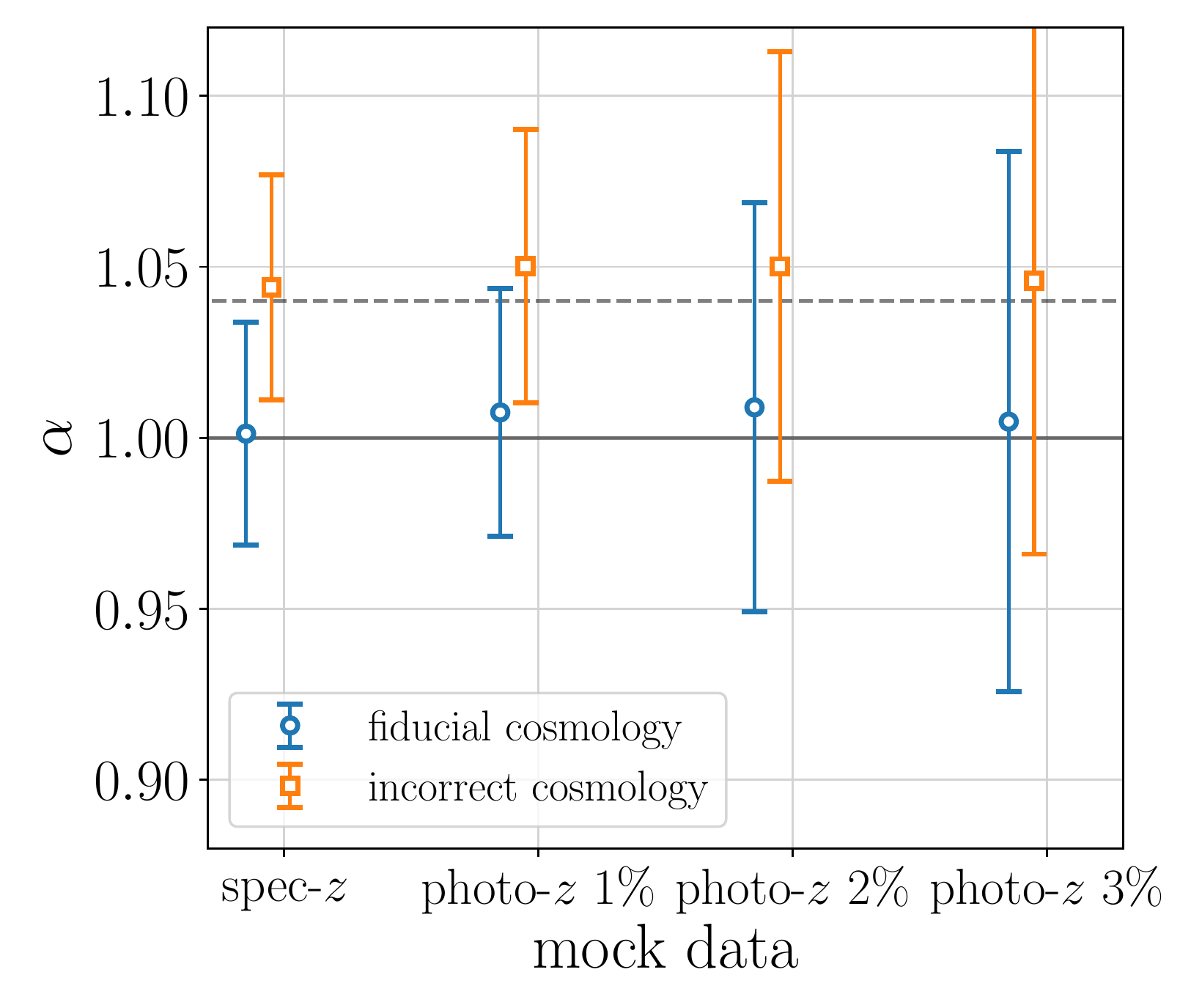}&
            \includegraphics[width=0.31\linewidth,keepaspectratio]{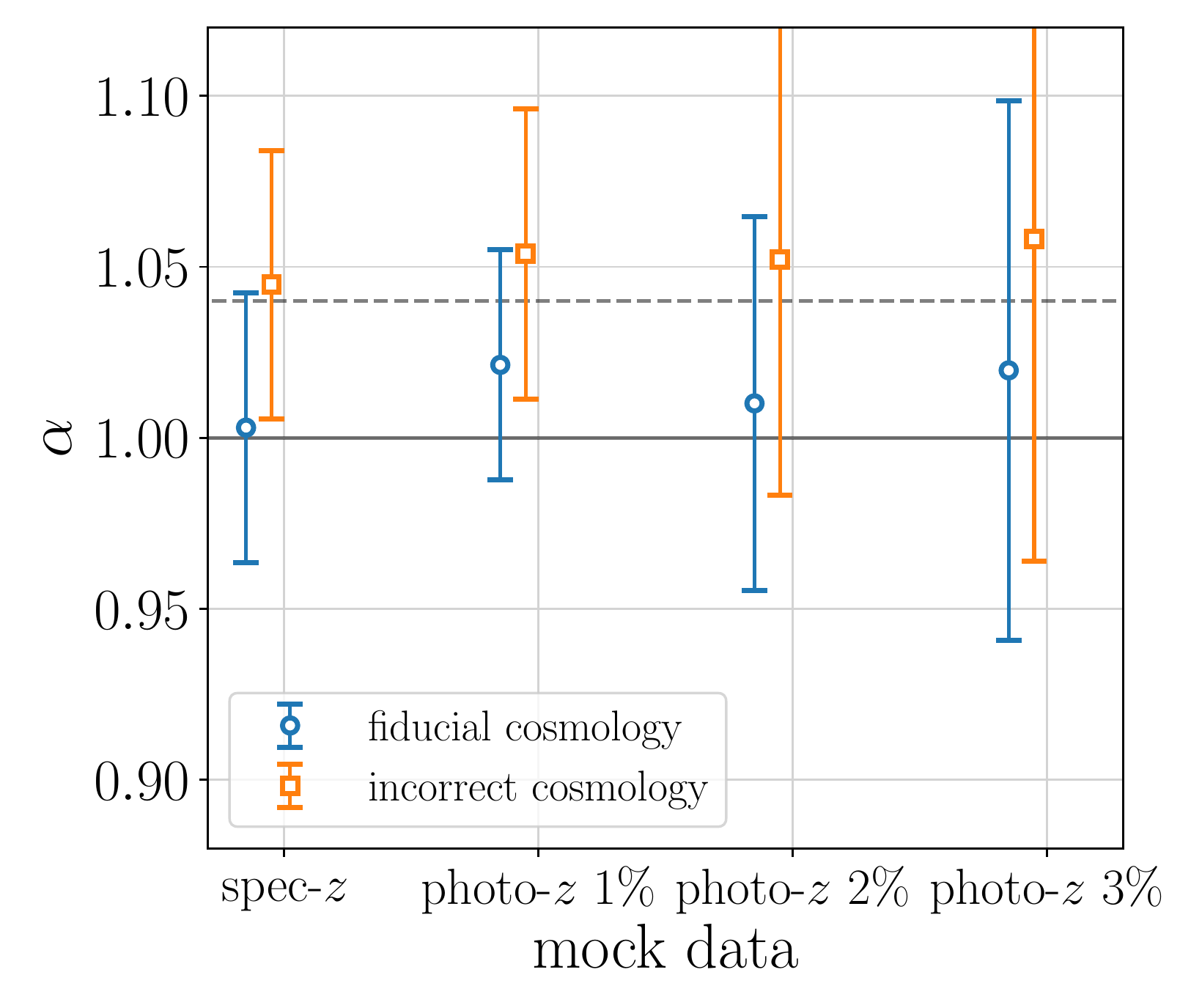}\\
            (a) $z$=0.251 &
            (b) $z$=0.617 &
            (c) $z$=1.03\\
        \end{tabular}
        \caption{
        The best-fit values of $\alpha$ at each redshift. From left to right, redshift is $z=0.251, 0.617$, and $1.03$ for fixed stellar mass limit samples being $M_* \ge 10^{11} M_{\sun} / h^2$ when the photo-$z$ error is known. The data in blue is the result when fiducial cosmology is used, and the data in orange is the result when incorrect cosmology is used. The expected value with the fiducial cosmology is $\alpha =1$, so the solid line is drawn. And also the expected value with the incorrect cosmology is theoretically $\alpha = 1.04$, so the dotted line is drawn. These error bars represent 95\% confidence intervals.
        }
        \label{fig:alpha_value_each_redshift_SML11_each_fit_inc}
    \end{center}
\end{figure*}

The discussion so far focuses on the BAO shift parameter, $\alpha$.
However, we now aim to connect $\alpha$ and its corresponding constraint on a cosmological parameter $\Omega_{\rm m0}$ through Eq.~\eqref{eq:alpha}.
In this analysis, we assume a 10\% smaller value of $\Omega_{\rm m0}$ compared to its correct value and repeat the analysis, and this 10\% decrease in $\Omega_{\rm m0}$ will result in a corresponding 4\% increase in the value of $\alpha$.
Fig.~\ref{fig:alpha_value_each_redshift_SML11_each_fit_inc} 
shows the comparisons of $\alpha$ assuming the correct or 10\% smaller value of $\Omega_{\rm m0}$, which has the stellar mass limit of $M_* \ge 10^{11} M_{\sun} /h^2$.

\begin{figure}
    \begin{center}
            \includegraphics[width=0.8\linewidth,keepaspectratio]{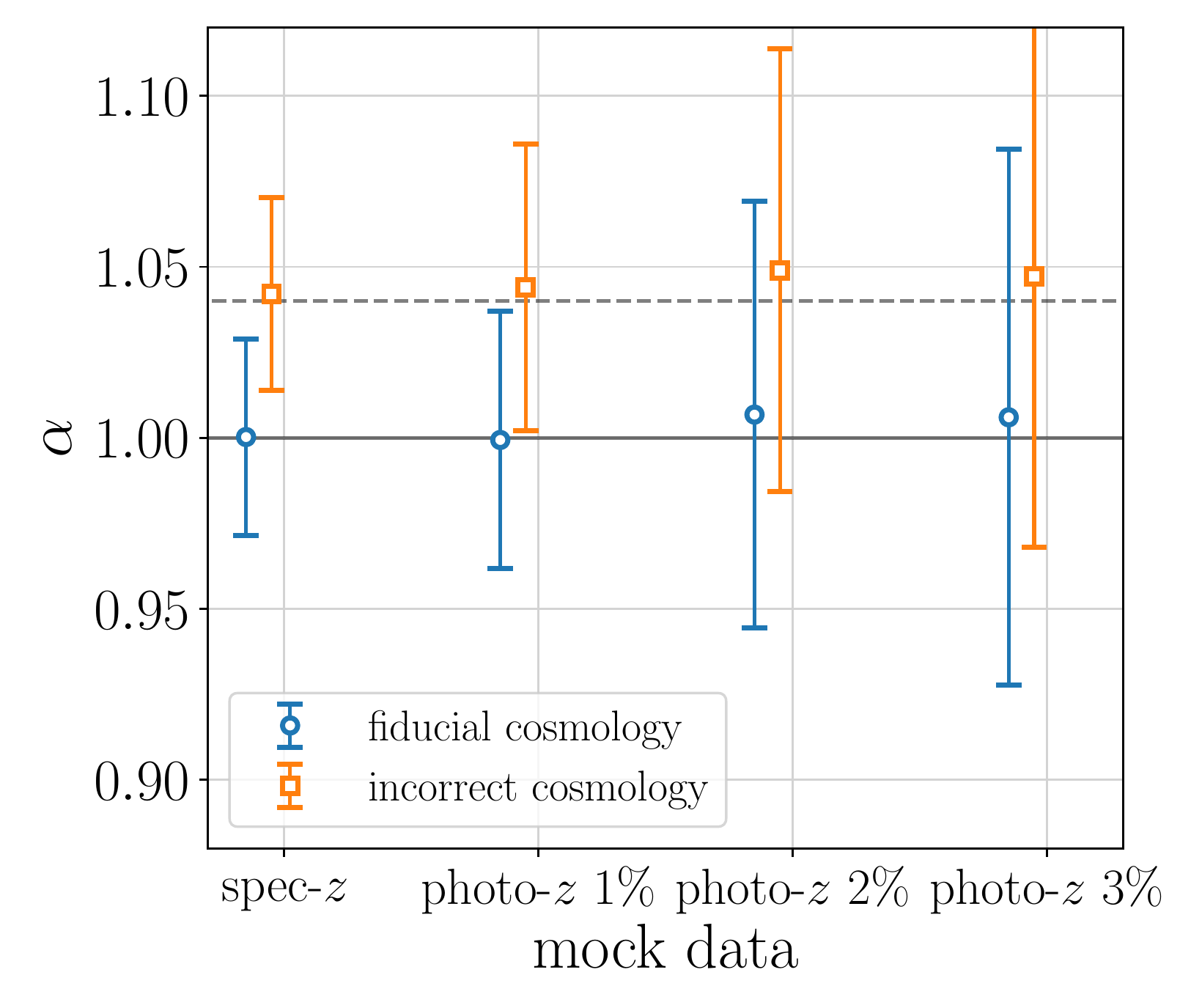}
        \caption{
        The result for the galaxy sample with $M_* \ge 10^{10.25} M_{\sun} / h^2$ at $z$=1.03. The best-fit values of $\alpha$ at $z=1.03$. The data in blue is the result when fiducial cosmology is assumed, and the data in orange is the result when incorrect cosmology is assumed. The expected value with the fiducial cosmology is $\alpha =1$ (solid line), while the expected value with the incorrect cosmology is $\alpha = 1.04$ (dotted line). These error bars represent 95\% confidence intervals.
        }
        \label{fig:alpha_value_each_redshift_SML10.25&cross_each_fit_inc}
    \end{center}
\end{figure}

When we assume a 10\% smaller value of $\Omega_{\rm m0}$, the expected value of $\alpha$ becomes 4\% larger. The best-fit value of $\alpha$ is consistent with the expected shift for all the cases we consider. This means that the BAO measurements using photometric galaxies can robustly constrain $\Omega_{\rm m0}$ even for the photo-$z$ uncertainty of 3\%.
If we focus on the photo-$z$ of $1\%$, the constraint on $\alpha$ for the case of a 10\% smaller value of $\Omega_{\rm m0}$ excludes the value of $\alpha=1$. 
This implies that the galaxy sample with the photo-$z$ uncertainty of 1\% with 1 [\gpc]$^3$ volume can constrain the $\Omega_{\rm m0}$ with $10\%$ accuracy.
We note that even at the highest redshift ($z$=1.03), the constraining power on $\Omega_{\rm m0}$ is not degraded, and we expect that combining the BAO measurement at different redshifts can further improve the constraining power.

Fig.~\ref{fig:alpha_value_each_redshift_SML10.25&cross_each_fit_inc} shows the result of the same experiment but for a denser sample (i.e. $M_* \ge 10^{10.25} M_{\sun} /h^2$ at $z=1.03$).
As is shown in Fig.~\ref{fig:alpha_value_each_redshift_SML10.25&cross_each_fit_inc}, the best-fit values of $\alpha$ are not biased. 
Furthermore, when we assume the 10\% smaller $\Omega_{\rm m0}$ for our fitting model, the best-fit values are unbiasedly constrained to the expected value of $\alpha = 1.04$.

\subsection{Prediction for LSST-like survey}
\label{ssec:LSST-like_covariance}
\begin{figure}
    \begin{center}
        \includegraphics[width=\linewidth,keepaspectratio]{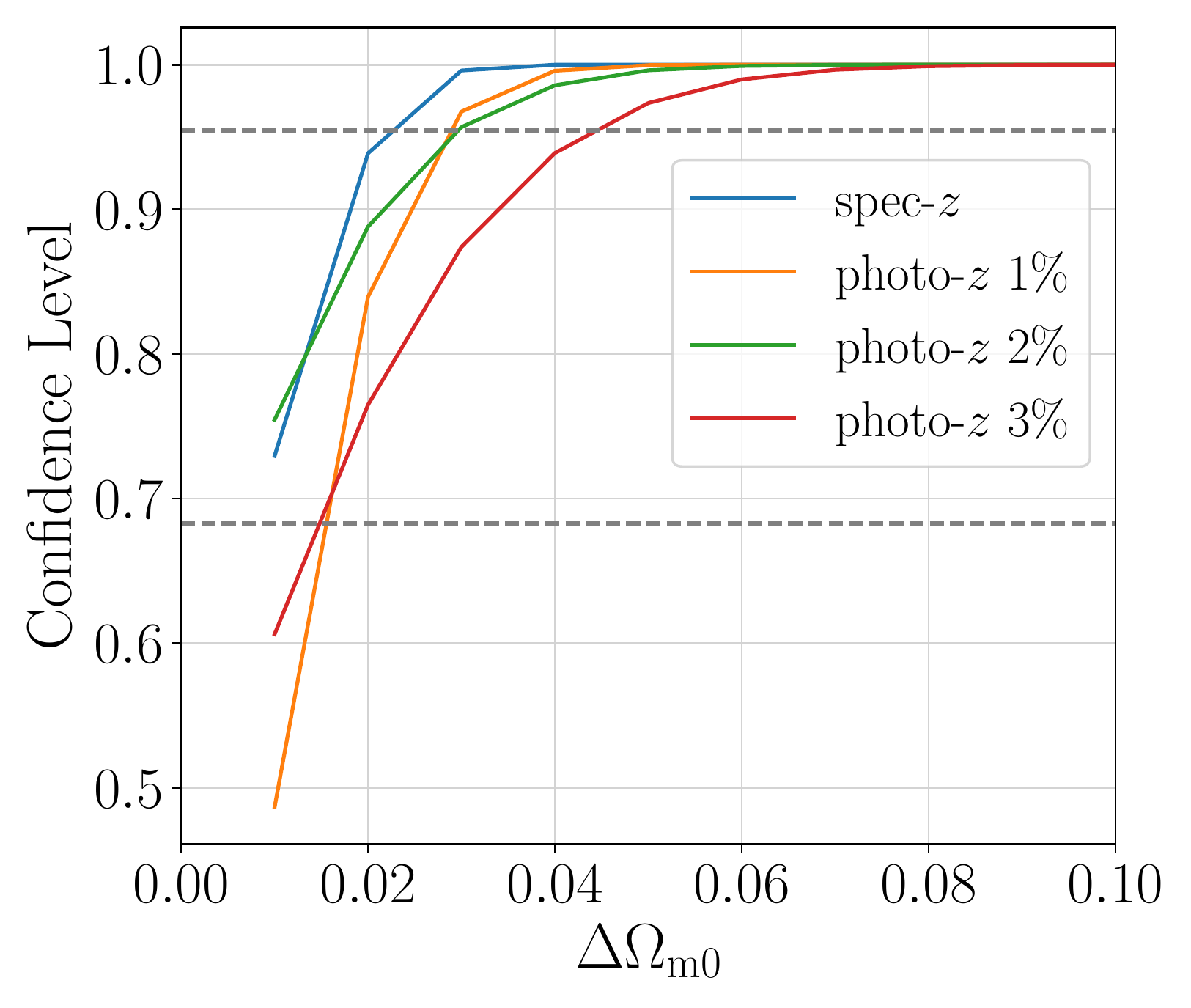}
        \caption{The best-fit values of $\alpha$ at $z$=1.03 ($M_* \ge 10^{10.25} M_{\sun}/h^2$) using LSST-like covariance. The horizontal axis shows a size of bias for the assumed value of $\Omega_{\rm m0}$ used in our fitting model with respect to fiducial cosmology. That is, $\Delta \Omega_{\rm m0}=0.00$ represents fiducial $\Omega_{\rm m0}$ of the mock galaxy sample. The vertical axis represents the confidence level at $\alpha =1$. The lower horizontal dotted line in grey is 1 $\sigma$ (68\%) limit, and the upper one is 2 $\sigma$ (95\%). Different colours correspond to spectroscopic galaxies (blue) and photometric galaxies with photo-$z$ error of 1\% (orange), 2\% (green), and 3\% (red). }
        \label{fig:DeltaOm_LSST}
    \end{center}
\end{figure}

We investigate how robustly LSST can constrain $\Omega_{\rm m0}$ using the BAO measurement with photo=$z$ uncertainties. To do that, we scale the covariance matrix to the LSST-like survey volume using Eq.~\eqref{eq:scaled_cov}.
Fig.~\ref{fig:DeltaOm_LSST} shows how well the BAO measurement can differentiate the correct $\Omega_{\rm m0}$ from the assumed cosmology used for the model template. For this analysis, we use the galaxy sample with $M_* \ge 10^{10.25} M_{\sun}/h^2$ at $z$=1.03.
We can constrain $\Delta \Omega_{\rm m0}<0.05$ with 95\% confidence region if photo-$z$ uncertainty is 3\%, and $\Delta \Omega_{\rm m0}<0.03$ for the case of 1\% photo-$z$ error.

\begin{table*}
  \begin{center}
      \caption{The fitting results in \S~\ref{ssec:incorrect_cosmology} when incorrect cosmology is assumed. 
      These rows correspond to each Section and $z=1.03$ ($M_* \ge 10^{10.25} M_{\sun}/h^2$) in \S~\ref{ssec:bao_measure_perror_isknown} is fiducial set-up in this paper.
      The cross-correlation is conducted between spec-$z$ and each photo-$z$ data, so the column of spec-$z$ data is blank.
      These columns correspond to the data which is used, and the column of spec-$z$ data is for comparison.
      These values are the best fits using the least $\chi^2$ fitting, and these errors are marginalised over the covariance of fitting parameters space generated after fitting.
      The error bars are with 95\% confidence.
      }
      \label{tab:fitting_results_in_Sec_incorrect}
      \begin{tabular}{c c c c c c c} \hline
        & stellar mass limit & spec-$z$ data & \multicolumn{3}{c}{photo-$z$ data} & ref. \\ \hline 
        & $M_* \ge 10^x M_{\sun}/h^2$ & & 1\% error & 2\% error & 3\% error & \\ \hline 
        $z=0.251$ & $x=11.00$ 
        & 1.041 $\pm$ 0.027 & 1.049 $\pm$ 0.035 & 1.057 $\pm$ 0.064 & 1.066 $\pm$ 0.077 & 
        \multirow{3}{*}{\S~\ref{ssec:bao_measure_perror_isknown}} \\
        $z=0.617$ & $x=11.00$ 
        & 1.044 $\pm$ 0.033 & 1.050 $\pm$ 0.040 & 1.050 $\pm$ 0.063 & 1.046 $\pm$ 0.080 & \\
        $z=1.03$  & $x=11.00$ 
        & 1.045 $\pm$ 0.039 & 1.054 $\pm$ 0.042 & 1.052 $\pm$ 0.069 & 1.058 $\pm$ 0.094 & \\ \hline \hline
        $z=1.03$  & $x=10.25$ 
        & 1.042 $\pm$ 0.028 & 1.044 $\pm$ 0.042 & 1.049 $\pm$ 0.065 & 1.047 $\pm$ 0.079 & 
        \S~\ref{ssec:bao_measure_perror_isknown}  \\ \hline \hline
        spec-$z$ fix template & $x=10.25$ 
        & 1.042 $\pm$ 0.028 & 1.029 $\pm$ 0.056 & 1.032 $\pm$ 0.088 & 1.040 $\pm$ 0.112 & 
        \multirow{3}{*}{\S~\ref{ssec:bao_measure_perror_isunknown}} \\
        photo-$z$ 1\% fix template & $x=10.25$ 
        & 1.064 $\pm$ 0.037 & 1.044 $\pm$ 0.042 & 1.068 $\pm$ 0.084 & 1.096 $\pm$ 0.133 & \\
        photo-$z$ 2\% fix template & $x=10.25$ 
        & 1.050 $\pm$ 0.028 & 1.020 $\pm$ 0.027 & 1.049 $\pm$ 0.065 & 1.050 $\pm$ 0.080 & \\ \hline
        \multirow{2}{*}{cross-correlation with spec-$z$} & $x=10.25$ (photo-$z$)
        & & \multirow{2}{*}{1.037 $\pm$ 0.031} & \multirow{2}{*}{1.053 $\pm$ 0.061} & \multirow{2}{*}{1.050 $\pm$ 0.077} & \multirow{2}{*}{\S~\ref{ssec:bao_measure_perror_isknown_cross}} \\
        & $x=11.00$ (spec-$z$) & \multicolumn{5}{c}{} \\ \hline
      \end{tabular}
  \end{center}
\end{table*}

\subsection{skewed photo-$z$ non-Gaussian distribution check}
\label{ssec:skewness}
\begin{figure*}
    \begin{center}
        \begin{tabular}{cc}
            \includegraphics[width=0.45\linewidth,keepaspectratio]{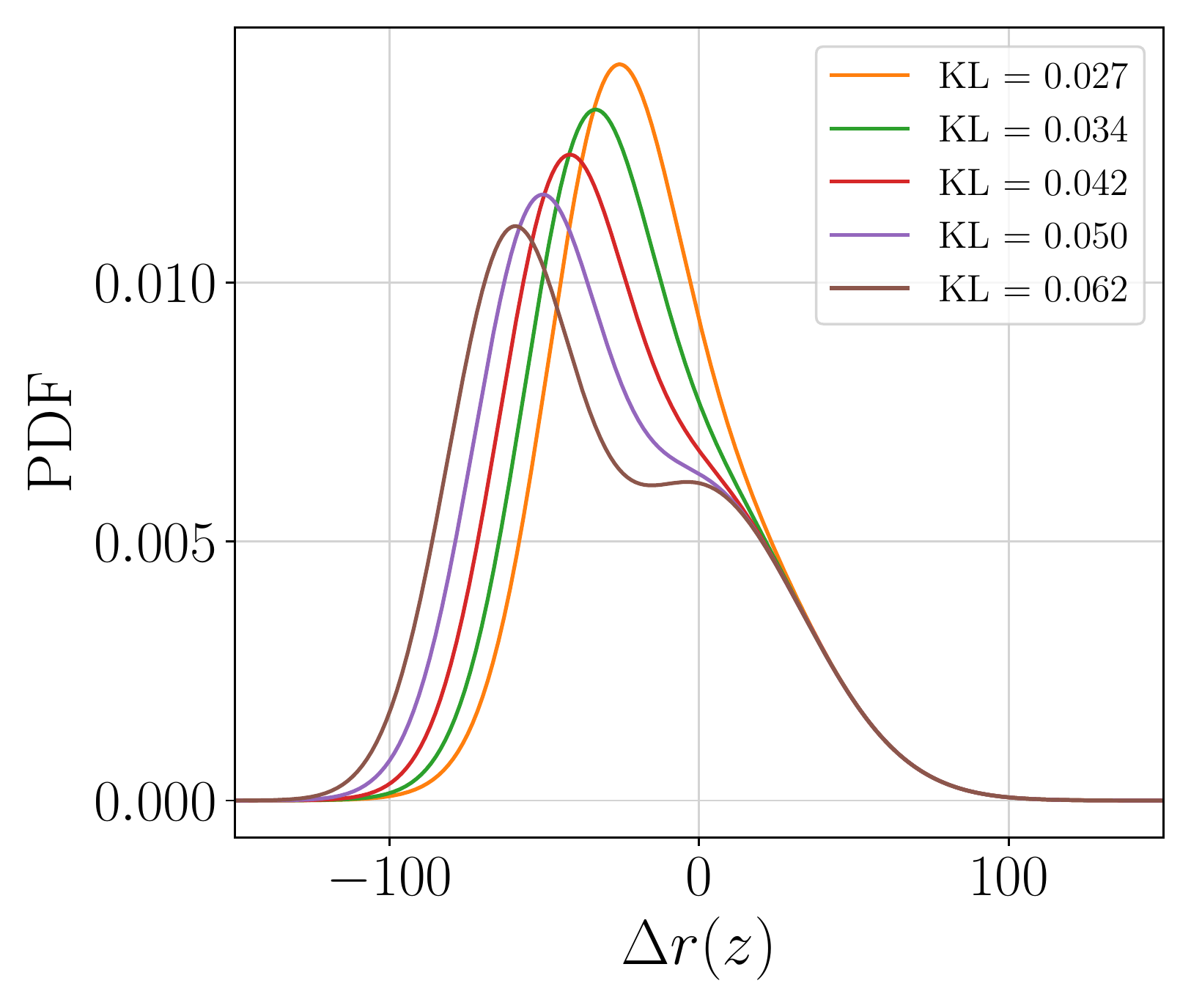}&
            \includegraphics[width=0.45\linewidth,keepaspectratio]{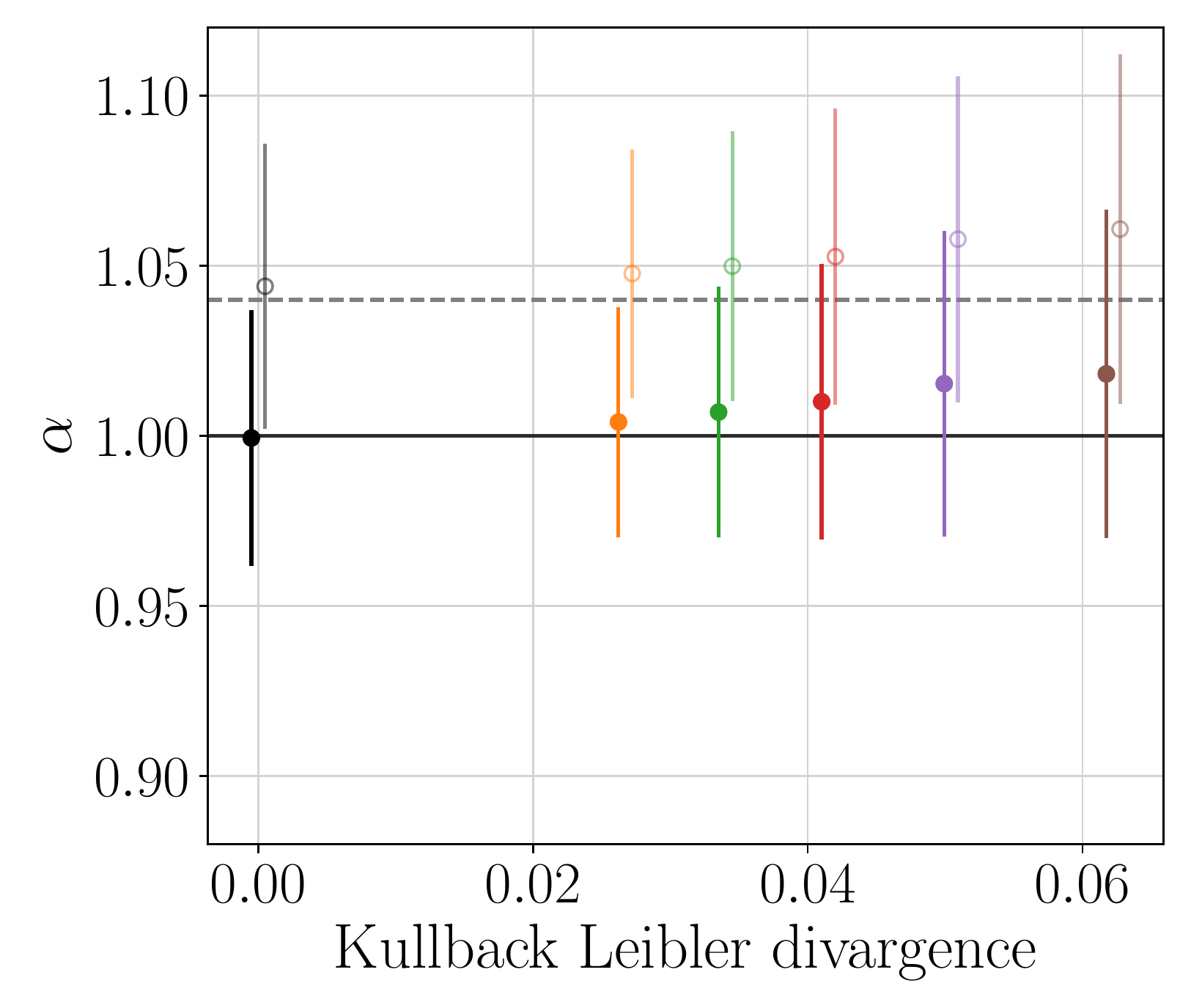}\\
            (a) skewed non-Gaussian photo-$z$ distribution &
            (b) BAO scale against KL divergence \\
        \end{tabular}
        \caption{The result of the effect of the photo-$z$ distribution is not Gaussian but skewed non-Gaussian. The skewed non-Gaussian photo-$z$ distributions (snG) produced by two Gaussian summations are used in the galaxy mock catalogue and fit them using Gaussian photo-$z$ distribution template whose mean and STD are the same as snG, that is when the mean and STD of photo-$z$ distribution could be reproduced. The usage guide in this figure represents the one varying mean of two Gaussian consisting of skewed non-Gaussian, and these colours correspond between this figure (a) and (b).
        (a) The skewed non-Gaussian photo-$z$ distributions by combining two Gaussian distributions where one mean of the two Gaussian is varied while the other is fixed to zero. The horizontal axis represents the comoving distance converted from photo-$z$ - spec-$z$, and the vertical axis is a probability distribution function of skewed non-Gaussian as a function of KL divergence.
        (b) The constraint on $\alpha$ for the case of the snG photo-$z$ distribution. The horizontal axis is the KL divergence between the skewed non-Gaussian and the Gaussian when the mean and the STD of skewed non-Gaussian are known. If KL divergence is 0, the photo-$z$ distribution is Gaussian with the photo-$z$ error of 1\%. These error bars are from the 112 mock data to fit and represent 95\% confidence intervals.
        The data in the filled circle is with fiducial cosmology, and the data in the open circle is with incorrect cosmology assuming 10\% smaller value of $\Omega_{\rm m0}$.
        }
        \label{fig:skewed_non-Gaussian}
     \end{center}
\end{figure*}

A photo-$z$ distribution is often assumed to be Gaussian. However, the photo-$z$ distribution does not need to follow Gaussian. To investigate whether the skewness of the photo-$z$ distribution biases the constraint on $\alpha$, we generate the mock galaxies following a non-Gaussian distribution. Specifically, we assume that the non-Gaussian distribution is a superposition of two different Gaussian distributions with the means $\mu=(0,\Delta \mu)$ and the standard deviations $\sigma=$(20, 33)\mpc\, where $\Delta \mu$ ranges from -30 \mpc\ to -62 \mpc. The amplitudes of two Gaussian distributions can be defined so that each of them is separately normalised to unity.
The mean of the composite distribution $\mu_{\rm snG}$ changes -15~\mpc\ to -31~\mpc\ depending on $\Delta \mu$ (see left panel of Fig.~\ref{fig:skewed_non-Gaussian}), and standard deviation $\sigma_{\rm snG}$ also changes 31 \mpc\ to 41 \mpc. We call this distribution a skewed non-Gaussian (snG) distribution. 
We note that even when the mean of snG distribution is shifted from zero, it does not affect the BAO measurement and thus $\alpha$ is not biased as long as the mean shift is common for all the galaxies because the two-point correlation function is a function of the separation of two galaxies. Instead, the standard deviation of snG distribution largely affects the constraint on $\alpha$. In practice, the galaxy photo-z pdf differs by galaxies but this is beyond the scope of this paper and we leave it for future work.
From our mock galaxy catalogue, we randomly fluctuate the line-of-sight position according to this snG distribution. However, we still compute our fitting model described in Eq.~\eqref{eq:xi_int} assuming the photo-$z$ distribution is Gaussian with mean zero. The scatter of the Gaussian is given by $\sigma_{\rm snG}$.
We quantify the disagreement of the full snG and the approximated Gaussian distributions with the correct standard deviation using the Kullback Leibler (KL) divergence.
The right panel of Fig.~\ref{fig:skewed_non-Gaussian} shows the constraints on $\alpha$ as a function of KL divergence corresponding to the photo-$z$ distribution shown in the left panel of Fig.~\ref{fig:skewed_non-Gaussian}. The best-fit value of $\alpha$ gets slightly biased as the KL divergence becomes larger. However, the shift from $\alpha=1$ is well within 2$\sigma$. 
This implies that the BAO measurement is robust against the skewness of the photo-$z$ distribution, and a priori knowledge of the scatter of the photo-$z$ uncertainty is sufficient for the unbiased BAO measurement.
In the case of the LSST-like survey, the error will become one-third, and thus the requirement will be further stringent. We see that the BAO measurement is unbiased when KL$<0.04$.
We verify this result consistent with \cite{Chaves-Montero+2018}.
However, we need to be careful with different approaches between ours and \cite{Chaves-Montero+2018}: They use multipole power spectra, sub-percent photo-$z$ error up to 1\% error, larger periodic boxes, and different definitions of skewed photo-$z$ distribution. 
In a more realistic situation of galaxy distribution, the photo-$z$ often has some population of catastrophic failure, called outlier. Given that $z=0.25$ galaxy sample contains 10\% of outliers at $z=3$ \citep{Xiao+2022}, the effect of outlier can be approximated by replacing the 10\% of galaxies with the random points. This simply decreases the amplitude of the clustering and the effect of the estimation of $\alpha$ is not largely affected.
    



\section{Summary}
\label{sec:summary}
In this paper, we explore the robustness of measuring the Baryon Acoustic Oscillation (BAO) peak, assuming the use of photometric galaxies, and quantify the accuracy of the BAO measurement that photometric redshift surveys can achieve.
We first construct mock galaxy catalogues considering the galaxies with $M_* \ge 10^{11} M_{\sun}/h^2$ and $M_* \ge 10^{10.25} M_{\sun}/h^2$ with various photo-$z$ uncertainties.

First, we assume that the photometric redshift distribution is Gaussian and the uncertainty ($\sigma_P$) is accurately estimated as described in  Eq.~\eqref{eq:xi_int}. 
We find that $\alpha$ is unbiasedly measured even in the case of photo-$z$ uncertainty of 3\% at $z=$0.251, 0.617, and 1.03, but the size of the error gets larger as photo-$z$ error increases from 4\% to 9\% corresponding to the photo-$z$ error of 1\% to 3\%.
We also find that a small galaxy number density can produce noise in the covariance matrix, and bias the measurement of $\alpha$ slightly. 
We generate a denser galaxy sample with  $M_* \ge 10^{10.25} M_{\sun}/h^2$ at $z=1.03$ and compare the result with the case of $M_* \ge 10^{10.25} M_{\sun}/h^2$.
We find that a higher number density can improve an unbiased measurement of $\alpha$ as shown in Fig.~\ref{fig:alpha_value_S004_compared_SML11}.

In \S~\ref{ssec:bao_measure_perror_isunknown}, we explore the case when we do not know the correct photo-$z$ uncertainty a priori or we either underestimate or overestimate it.
To test this, we use the fitting model assuming the wrong photo-$z$ uncertainty and repeat the analysis.
As shown in Fig.~\ref{fig:alpha_value_S004_fixed_photoz_fit}, underestimating the photo-$z$ error causes a larger error of $\alpha$.
Conversely, overestimating the photo-$z$ error can bias the constraint on $\alpha$.

Additionally, we measure cross-correlation functions between a sparse spectroscopic redshift sample and a dense photometric redshift sample at $z=1.03$.
We use $M_* \ge 10^{11} M_{\sun}/h^2$ sample for spec-$z$ and $M_* \ge 10^{10.25} M_{\sun}/h^2$ sample for photo-$z$.
Fig.~\ref{fig:alpha_value_S004_cross} shows that the constraint on $\alpha$ gets improved by roughly 30\% compared to the case of the auto-correlation functions measured from photometric redshift galaxies.

We also discuss whether the BAO peak is shifted expectedly by changing the assumed cosmology used for the template function from the fiducial cosmology.
To test this, we use a 10\% smaller value of $\Omega_{\rm m0}$ and repeat the analysis. Fig.~\ref{fig:alpha_value_each_redshift_SML11_each_fit_inc} shows that the best-fit value of $\alpha$ is shifted in an expected manner even for the case of photo-$z$ error of 3\%.
Additionally, we estimate the constraining power of $\alpha$ and its robustness for the  LSST-like survey assuming 11 [\gpc]$^3$ survey volume at $z=1.03$ ($M_* \ge 10^{10.25} M_{\sun}/h^2$). 
We find that $\Omega_{\rm m0}$ can be constrained in 95\% confidence level $\sigma(\Omega_{\rm m0}) = 0.03 (0.05)$ for the case of the photo-$z$ error of 1\% (3\%).

At last, we explore the case that the photo-$z$ uncertainty distribution is skewed non-Gaussian.
We make the skewed non-Gaussian photo-$z$ distribution (snG) using the superposition of two Gaussian distributions.
We conclude that the detailed shape difference has little effect on the measurement of the BAO, and the most important factor in obtaining an unbiased estimate of $\alpha$ parameter is an accurate estimate of the scatter of the distribution. In other words, prior knowledge of the photo-$z$ uncertainty is sufficient for the robust photo-$z$ BAO analysis.

\section*{Acknowledgements}
We thank Shogo Ishikawa, Masamune Oguri, Teppei Okumura, Masahiro Takada, and Atsushi Taruya for useful discussions. 
This research is supported by a grant from the Hayakawa Satio Fund awarded by the Astronomical Society of Japan.
This paper is partly supported by JSPS KAKENHI Grant Numbers JP19H00677, JP20H01932, JP20H05861, JP20H05855, JP21H01081, JP22K03634, JP22K21349, and JP23H00108.

\section*{data availability}
The mock galaxy data in this article will be shared upon a reasonable request to the corresponding author.


\bibliographystyle{mnras}
\bibliography{main} 

\begin{thebibliography}{}
\makeatletter
\relax
\def\mn@urlcharsother{\let\do\@makeother \do\$\do\&\do\#\do\^\do\_\do\%\do\~}
\def\mn@doi{\begingroup\mn@urlcharsother \@ifnextchar [ {\mn@doi@}
  {\mn@doi@[]}}
\def\mn@doi@[#1]#2{\def\@tempa{#1}\ifx\@tempa\@empty \href
  {http://dx.doi.org/#2} {doi:#2}\else \href {http://dx.doi.org/#2} {#1}\fi
  \endgroup}
\def\mn@eprint#1#2{\mn@eprint@#1:#2::\@nil}
\def\mn@eprint@arXiv#1{\href {http://arxiv.org/abs/#1} {{\tt arXiv:#1}}}
\def\mn@eprint@dblp#1{\href {http://dblp.uni-trier.de/rec/bibtex/#1.xml}
  {dblp:#1}}
\def\mn@eprint@#1:#2:#3:#4\@nil{\def\@tempa {#1}\def\@tempb {#2}\def\@tempc
  {#3}\ifx \@tempc \@empty \let \@tempc \@tempb \let \@tempb \@tempa \fi \ifx
  \@tempb \@empty \def\@tempb {arXiv}\fi \@ifundefined
  {mn@eprint@\@tempb}{\@tempb:\@tempc}{\expandafter \expandafter \csname
  mn@eprint@\@tempb\endcsname \expandafter{\@tempc}}}

\bibitem[\protect\citeauthoryear{{Abbott} et~al.,}{{Abbott}
  et~al.}{2019}]{Abbott+2019}
{Abbott} T.~M.~C.,  et~al., 2019, \mn@doi [\mnras] {10.1093/mnras/sty3351},
  \href {https://ui.adsabs.harvard.edu/abs/2019MNRAS.483.4866A} {483, 4866}

\bibitem[\protect\citeauthoryear{{Abbott} et~al.,}{{Abbott}
  et~al.}{2022}]{Abbott+2022}
{Abbott} T.~M.~C.,  et~al., 2022, \mn@doi [\prd] {10.1103/PhysRevD.105.043512},
  \href {https://ui.adsabs.harvard.edu/abs/2022PhRvD.105d3512A} {105, 043512}

\bibitem[\protect\citeauthoryear{{Aihara} et~al.,}{{Aihara}
  et~al.}{2018}]{Aihara+2018}
{Aihara} H.,  et~al., 2018, \mn@doi [\pasj] {10.1093/pasj/psx066}, \href
  {https://ui.adsabs.harvard.edu/abs/2018PASJ...70S...4A} {70, S4}

\bibitem[\protect\citeauthoryear{{Amendola} et~al.,}{{Amendola}
  et~al.}{2018}]{euclid2018}
{Amendola} L.,  et~al., 2018, \mn@doi [Living Reviews in Relativity]
  {10.1007/s41114-017-0010-3}, \href
  {https://ui.adsabs.harvard.edu/abs/2018LRR....21....2A} {21, 2}

\bibitem[\protect\citeauthoryear{{Behroozi}, {Wechsler}  \& {Wu}}{{Behroozi}
  et~al.}{2013}]{Behroozi+2013}
{Behroozi} P.~S.,  {Wechsler} R.~H.,   {Wu} H.-Y.,  2013, \mn@doi [\apj]
  {10.1088/0004-637X/762/2/109}, \href
  {https://ui.adsabs.harvard.edu/abs/2013ApJ...762..109B} {762, 109}

\bibitem[\protect\citeauthoryear{{Carroll}, {Press}  \& {Turner}}{{Carroll}
  et~al.}{1992}]{Carroll+1992}
{Carroll} S.~M.,  {Press} W.~H.,   {Turner} E.~L.,  1992, \mn@doi [\araa]
  {10.1146/annurev.aa.30.090192.002435}, \href
  {https://ui.adsabs.harvard.edu/abs/1992ARA&A..30..499C} {30, 499}

\bibitem[\protect\citeauthoryear{{Chan} et~al.,}{{Chan}
  et~al.}{2022a}]{Chan+2022a}
{Chan} K.~C.,  et~al., 2022a, \mn@doi [\prd] {10.1103/PhysRevD.106.123502},
  \href {https://ui.adsabs.harvard.edu/abs/2022PhRvD.106l3502C} {106, 123502}

\bibitem[\protect\citeauthoryear{{Chan}, {Ferrero}, {Avila}, {Ross}, {Crocce}
  \& {Gazta{\~n}aga}}{{Chan} et~al.}{2022b}]{Chan+2022b}
{Chan} K.~C.,  {Ferrero} I.,  {Avila} S.,  {Ross} A.~J.,  {Crocce} M.,
  {Gazta{\~n}aga} E.,  2022b, \mn@doi [\mnras] {10.1093/mnras/stac340}, \href
  {https://ui.adsabs.harvard.edu/abs/2022MNRAS.511.3965C} {511, 3965}

\bibitem[\protect\citeauthoryear{{Chaves-Montero}, {Angulo}  \&
  {Hern{\'a}ndez-Monteagudo}}{{Chaves-Montero}
  et~al.}{2018}]{Chaves-Montero+2018}
{Chaves-Montero} J.,  {Angulo} R.~E.,   {Hern{\'a}ndez-Monteagudo} C.,  2018,
  \mn@doi [\mnras] {10.1093/mnras/sty924}, \href
  {https://ui.adsabs.harvard.edu/abs/2018MNRAS.477.3892C} {477, 3892}

\bibitem[\protect\citeauthoryear{{Crocce}, {Pueblas}  \&
  {Scoccimarro}}{{Crocce} et~al.}{2006}]{Crocce+2006}
{Crocce} M.,  {Pueblas} S.,   {Scoccimarro} R.,  2006, \mn@doi [\mnras]
  {10.1111/j.1365-2966.2006.11040.x}, \href
  {https://ui.adsabs.harvard.edu/abs/2006MNRAS.373..369C} {373, 369}

\bibitem[\protect\citeauthoryear{{Dore} et~al.,}{{Dore}
  et~al.}{2019}]{WFIRST2019}
{Dore} O.,  et~al., 2019, \baas, \href
  {https://ui.adsabs.harvard.edu/abs/2019BAAS...51c.341D} {51, 341}

\bibitem[\protect\citeauthoryear{{Eisenstein} \& {Hu}}{{Eisenstein} \&
  {Hu}}{1998}]{Eisenstein_Hu1998}
{Eisenstein} D.~J.,  {Hu} W.,  1998, \mn@doi [\apj] {10.1086/305424}, \href
  {https://ui.adsabs.harvard.edu/abs/1998ApJ...496..605E} {496, 605}

\bibitem[\protect\citeauthoryear{{Eisenstein} et~al.,}{{Eisenstein}
  et~al.}{2005}]{Eisenstein+2005}
{Eisenstein} D.~J.,  et~al., 2005, \mn@doi [\apj] {10.1086/466512}, \href
  {https://ui.adsabs.harvard.edu/abs/2005ApJ...633..560E} {633, 560}

\bibitem[\protect\citeauthoryear{{Eisenstein}, {Seo}  \& {White}}{{Eisenstein}
  et~al.}{2007}]{Eisenstein+2007}
{Eisenstein} D.~J.,  {Seo} H.-J.,   {White} M.,  2007, \mn@doi [\apj]
  {10.1086/518755}, \href
  {https://ui.adsabs.harvard.edu/abs/2007ApJ...664..660E} {664, 660}

\bibitem[\protect\citeauthoryear{{Fang}, {Eifler}, {Schaan}, {Huang}, {Krause}
  \& {Ferraro}}{{Fang} et~al.}{2022}]{Xiao+2022}
{Fang} X.,  {Eifler} T.,  {Schaan} E.,  {Huang} H.-J.,  {Krause} E.,
  {Ferraro} S.,  2022, \mn@doi [\mnras] {10.1093/mnras/stab3410}, \href
  {https://ui.adsabs.harvard.edu/abs/2022MNRAS.509.5721F} {509, 5721}

\bibitem[\protect\citeauthoryear{{Groth} \& {Peebles}}{{Groth} \&
  {Peebles}}{1977}]{Groth&Peebles1977}
{Groth} E.~J.,  {Peebles} P.~J.~E.,  1977, \mn@doi [\apj] {10.1086/155588},
  \href {https://ui.adsabs.harvard.edu/abs/1977ApJ...217..385G} {217, 385}

\bibitem[\protect\citeauthoryear{{Hartlap}, {Simon}  \& {Schneider}}{{Hartlap}
  et~al.}{2007}]{Hartlap+2007}
{Hartlap} J.,  {Simon} P.,   {Schneider} P.,  2007, \mn@doi [\aap]
  {10.1051/0004-6361:20066170}, \href
  {https://ui.adsabs.harvard.edu/abs/2007A&A...464..399H} {464, 399}

\bibitem[\protect\citeauthoryear{{Ho} et~al.,}{{Ho} et~al.}{2012}]{Ho:2012}
{Ho} S.,  et~al., 2012, \mn@doi [\apj] {10.1088/0004-637X/761/1/14}, \href
  {https://ui.adsabs.harvard.edu/abs/2012ApJ...761...14H} {761, 14}

\bibitem[\protect\citeauthoryear{{H{\"u}tsi}}{{H{\"u}tsi}}{2010}]{Hutsi2010}
{H{\"u}tsi} G.,  2010, \mn@doi [\mnras] {10.1111/j.1365-2966.2009.15824.x},
  \href {https://ui.adsabs.harvard.edu/abs/2010MNRAS.401.2477H} {401, 2477}

\bibitem[\protect\citeauthoryear{{Ishikawa}, {Okumura}, {Oguri}  \&
  {Lin}}{{Ishikawa} et~al.}{2021}]{ishikawa+2021}
{Ishikawa} S.,  {Okumura} T.,  {Oguri} M.,   {Lin} S.-C.,  2021, \mn@doi [\apj]
  {10.3847/1538-4357/ac1f90}, \href
  {https://ui.adsabs.harvard.edu/abs/2021ApJ...922...23I} {922, 23}

\bibitem[\protect\citeauthoryear{{Kaiser}}{{Kaiser}}{1987}]{Kaiser1987}
{Kaiser} N.,  1987, \mn@doi [\mnras] {10.1093/mnras/227.1.1}, \href
  {https://ui.adsabs.harvard.edu/abs/1987MNRAS.227....1K} {227, 1}

\bibitem[\protect\citeauthoryear{{Kuijken} et~al.,}{{Kuijken}
  et~al.}{2015}]{KiDs2015}
{Kuijken} K.,  et~al., 2015, \mn@doi [\mnras] {10.1093/mnras/stv2140}, \href
  {https://ui.adsabs.harvard.edu/abs/2015MNRAS.454.3500K} {454, 3500}

\bibitem[\protect\citeauthoryear{{LSST Science Collaboration} et~al.,}{{LSST
  Science Collaboration} et~al.}{2009}]{LSST2009}
{LSST Science Collaboration} et~al., 2009, arXiv e-prints, \href
  {https://ui.adsabs.harvard.edu/abs/2009arXiv0912.0201L} {p. arXiv:0912.0201}

\bibitem[\protect\citeauthoryear{{Landy} \& {Szalay}}{{Landy} \&
  {Szalay}}{1993}]{Landy_Szalay1993}
{Landy} S.~D.,  {Szalay} A.~S.,  1993, \mn@doi [\apj] {10.1086/172900}, \href
  {https://ui.adsabs.harvard.edu/abs/1993ApJ...412...64L} {412, 64}

\bibitem[\protect\citeauthoryear{{Lesgourgues} \& {Tram}}{{Lesgourgues} \&
  {Tram}}{2011}]{CLASS2011}
{Lesgourgues} J.,  {Tram} T.,  2011, \mn@doi [\jcap]
  {10.1088/1475-7516/2011/09/032}, \href
  {https://ui.adsabs.harvard.edu/abs/2011JCAP...09..032L} {2011, 032}

\bibitem[\protect\citeauthoryear{{Mehta}, {Seo}, {Eckel}, {Eisenstein},
  {Metchnik}, {Pinto}  \& {Xu}}{{Mehta} et~al.}{2011}]{Mehta_2011}
{Mehta} K.~T.,  {Seo} H.-J.,  {Eckel} J.,  {Eisenstein} D.~J.,  {Metchnik} M.,
  {Pinto} P.,   {Xu} X.,  2011, \mn@doi [\apj] {10.1088/0004-637X/734/2/94},
  \href {https://ui.adsabs.harvard.edu/abs/2011ApJ...734...94M} {734, 94}

\bibitem[\protect\citeauthoryear{{Nishimichi} et~al.,}{{Nishimichi}
  et~al.}{2019}]{Nishimichi+2019}
{Nishimichi} T.,  et~al., 2019, \mn@doi [\apj] {10.3847/1538-4357/ab3719},
  \href {https://ui.adsabs.harvard.edu/abs/2019ApJ...884...29N} {884, 29}

\bibitem[\protect\citeauthoryear{{Nishizawa}, {Oguri}  \& {Takada}}{{Nishizawa}
  et~al.}{2013}]{Nishizawa+2013}
{Nishizawa} A.~J.,  {Oguri} M.,   {Takada} M.,  2013, \mn@doi [\mnras]
  {10.1093/mnras/stt761}, \href
  {https://ui.adsabs.harvard.edu/abs/2013MNRAS.433..730N} {433, 730}

\bibitem[\protect\citeauthoryear{{Oguri} et~al.,}{{Oguri}
  et~al.}{2018}]{Oguri+2018}
{Oguri} M.,  et~al., 2018, \mn@doi [\pasj] {10.1093/pasj/psx042}, \href
  {https://ui.adsabs.harvard.edu/abs/2018PASJ...70S..20O} {70, S20}

\bibitem[\protect\citeauthoryear{{Padmanabhan} \& {White}}{{Padmanabhan} \&
  {White}}{2009}]{PadmanabhanWhite_2009}
{Padmanabhan} N.,  {White} M.,  2009, \mn@doi [\prd]
  {10.1103/PhysRevD.80.063508}, \href
  {https://ui.adsabs.harvard.edu/abs/2009PhRvD..80f3508P} {80, 063508}

\bibitem[\protect\citeauthoryear{{Patej} \& {Eisenstein}}{{Patej} \&
  {Eisenstein}}{2018}]{Patej+2018}
{Patej} A.,  {Eisenstein} D.~J.,  2018, \mn@doi [\mnras]
  {10.1093/mnras/sty870}, \href
  {https://ui.adsabs.harvard.edu/abs/2018MNRAS.477.5090P} {477, 5090}

\bibitem[\protect\citeauthoryear{{Planck Collaboration} et~al.,}{{Planck
  Collaboration} et~al.}{2016}]{Planck2016}
{Planck Collaboration} et~al., 2016, \mn@doi [\aap]
  {10.1051/0004-6361/201525830}, \href
  {https://ui.adsabs.harvard.edu/abs/2016A&A...594A..13P} {594, A13}

\bibitem[\protect\citeauthoryear{{Ross} et~al.,}{{Ross}
  et~al.}{2017}]{Ross+2017b}
{Ross} A.~J.,  et~al., 2017, \mn@doi [\mnras] {10.1093/mnras/stx2120}, \href
  {https://ui.adsabs.harvard.edu/abs/2017MNRAS.472.4456R} {472, 4456}

\bibitem[\protect\citeauthoryear{{Scoccimarro}}{{Scoccimarro}}{1998}]{Scoccimarro1998}
{Scoccimarro} R.,  1998, \mn@doi [\mnras] {10.1046/j.1365-8711.1998.01845.x},
  \href {https://ui.adsabs.harvard.edu/abs/1998MNRAS.299.1097S} {299, 1097}

\bibitem[\protect\citeauthoryear{{Seo} \& {Eisenstein}}{{Seo} \&
  {Eisenstein}}{2005}]{Seo_2005}
{Seo} H.-J.,  {Eisenstein} D.~J.,  2005, \mn@doi [\apj] {10.1086/491599}, \href
  {https://ui.adsabs.harvard.edu/abs/2005ApJ...633..575S} {633, 575}

\bibitem[\protect\citeauthoryear{{Seo} et~al.,}{{Seo} et~al.}{2012}]{Seo+2012}
{Seo} H.-J.,  et~al., 2012, \mn@doi [\apj] {10.1088/0004-637X/761/1/13}, \href
  {https://ui.adsabs.harvard.edu/abs/2012ApJ...761...13S} {761, 13}

\bibitem[\protect\citeauthoryear{{Springel}}{{Springel}}{2005}]{Springel2005}
{Springel} V.,  2005, \mn@doi [\mnras] {10.1111/j.1365-2966.2005.09655.x},
  \href {https://ui.adsabs.harvard.edu/abs/2005MNRAS.364.1105S} {364, 1105}

\bibitem[\protect\citeauthoryear{{Springel}, {White}, {Tormen}  \&
  {Kauffmann}}{{Springel} et~al.}{2001}]{Springel+2001}
{Springel} V.,  {White} S. D.~M.,  {Tormen} G.,   {Kauffmann} G.,  2001,
  \mn@doi [\mnras] {10.1046/j.1365-8711.2001.04912.x}, \href
  {https://ui.adsabs.harvard.edu/abs/2001MNRAS.328..726S} {328, 726}

\bibitem[\protect\citeauthoryear{{Taruya}, {Bernardeau}, {Nishimichi}  \&
  {Codis}}{{Taruya} et~al.}{2012}]{TaruyaBernardeau:2012}
{Taruya} A.,  {Bernardeau} F.,  {Nishimichi} T.,   {Codis} S.,  2012, \mn@doi
  [\prd] {10.1103/PhysRevD.86.103528}, \href
  {https://ui.adsabs.harvard.edu/abs/2012PhRvD..86j3528T} {86, 103528}

\bibitem[\protect\citeauthoryear{{The Dark Energy Survey Collaboration}}{{The
  Dark Energy Survey Collaboration}}{2005}]{DES2005}
{The Dark Energy Survey Collaboration} 2005, arXiv e-prints, \href
  {https://ui.adsabs.harvard.edu/abs/2005astro.ph.10346T} {pp
  astro--ph/0510346}

\bibitem[\protect\citeauthoryear{{Totsuji} \& {Kihara}}{{Totsuji} \&
  {Kihara}}{1969}]{Totsuji&Kihara1969}
{Totsuji} H.,  {Kihara} T.,  1969, \pasj, \href
  {https://ui.adsabs.harvard.edu/abs/1969PASJ...21..221T} {21, 221}

\bibitem[\protect\citeauthoryear{{Weinberg}, {Mortonson}, {Eisenstein},
  {Hirata}, {Riess}  \& {Rozo}}{{Weinberg} et~al.}{2013}]{Weinberg+2013}
{Weinberg} D.~H.,  {Mortonson} M.~J.,  {Eisenstein} D.~J.,  {Hirata} C.,
  {Riess} A.~G.,   {Rozo} E.,  2013, \mn@doi [\physrep]
  {10.1016/j.physrep.2013.05.001}, \href
  {https://ui.adsabs.harvard.edu/abs/2013PhR...530...87W} {530, 87}

\bibitem[\protect\citeauthoryear{{Xu}, {Padmanabhan}, {Eisenstein}, {Mehta}  \&
  {Cuesta}}{{Xu} et~al.}{2012}]{Xu+2012}
{Xu} X.,  {Padmanabhan} N.,  {Eisenstein} D.~J.,  {Mehta} K.~T.,   {Cuesta}
  A.~J.,  2012, \mn@doi [\mnras] {10.1111/j.1365-2966.2012.21573.x}, \href
  {https://ui.adsabs.harvard.edu/abs/2012MNRAS.427.2146X} {427, 2146}

\bibitem[\protect\citeauthoryear{{Zarrouk} et~al.,}{{Zarrouk}
  et~al.}{2021}]{Zarrouk+2021}
{Zarrouk} P.,  et~al., 2021, \mn@doi [\mnras] {10.1093/mnras/stab298}, \href
  {https://ui.adsabs.harvard.edu/abs/2021MNRAS.503.2562Z} {503, 2562}

\bibitem[\protect\citeauthoryear{{Zheng} et~al.,}{{Zheng}
  et~al.}{2005}]{Zheng+2005}
{Zheng} Z.,  et~al., 2005, \mn@doi [\apj] {10.1086/466510}, \href
  {https://ui.adsabs.harvard.edu/abs/2005ApJ...633..791Z} {633, 791}

\makeatother
\end{thebibliography}







\bsp	
\label{lastpage}
\end{document}